\documentclass[fleqn,usenatbib,useAMS]{mnras}

\usepackage{graphicx}	
\usepackage{amsmath}	
\usepackage{amssymb}	
\usepackage{multicol}   
\usepackage{bm}		    
\usepackage{pdflscape}	
\usepackage{ulem}
\usepackage{hyperref}
\usepackage{xcolor}

\newcommand{\je}[1]{#1}

\newcommand{\Ra}{$0.5 \times R_{\mathrm{200c}}$}

\newcommand{\Rc}{$1.00 \times R_{\mathrm{200c}}$}

\newcommand{\Rcritical}{$R_{\mathrm{200c}}$}
\newcommand{\mustar}{$\rm \mu_{\star}$}
\newcommand{\mustarTrue}{$\rm \mu_{\star,true}$}

\newcommand{\Mo}{\mathrm{M_{\odot}}}

\DeclareMathOperator\arctanh{arctanh}
\DeclareMathOperator{\erf}{erf} 

\usepackage{eso-pic}

\AddToShipoutPictureBG*{%
  \AtPageUpperLeft{%
    \hspace{0.75\paperwidth}%
    \raisebox{-3.5\baselineskip}{%
      \makebox[0pt][l]{\textnormal{DES-2023-0754}}
}}}%

\AddToShipoutPictureBG*{%
  \AtPageUpperLeft{%
    \hspace{0.75\paperwidth}%
    \raisebox{-4.5\baselineskip}{%
      \makebox[0pt][l]{\textnormal{FERMILAB-PUB-24-0022-PPD}}
}}}%

\usepackage[T1]{fontenc}
\usepackage{ae,aecompl}
\usepackage[mathlines]{lineno}

\usepackage{newtxtext,newtxmath}

\title[Copacabana]{Copacabana: A Probabilistic Membership Assignment Method for Galaxy Clusters}

\author[Esteves et al.]{
\parbox{\textwidth}{
\Large
J. ~H. ~Esteves$^{1}$\thanks{Contact e-mail: \href{mailto:jesteves@umich.edu}{jesteves@umich.edu}},
M.~E.~S.~Pereira,$^{2}$
M.~Soares-Santos,$^{1}$
J.~Annis,$^{3}$
A.~Farahi,$^{4}$
F.~Andrade-Oliveira,$^{1}$
P.~Barchi,$^{5}$
A.~Palmese,$^{6}$
H.~Lin,$^{3}$
B.~Welch,$^{7,8}$
H.-Y.~Wu,$^{9}$
M.~Aguena,$^{10}$
O.~Alves,$^{1}$
D.~Bacon,$^{11}$
S.~Bocquet,$^{12}$
D.~Brooks,$^{13}$
A.~Carnero~Rosell,$^{14,10,15}$
J.~Carretero,$^{16}$
M.~Costanzi,$^{17,18,19}$
L.~N.~da Costa,$^{10}$
J.~De~Vicente,$^{20}$
P.~Doel,$^{13}$
S.~Everett,$^{21}$
B.~Flaugher,$^{3}$
J.~Frieman,$^{3,22}$
J.~Garc\'ia-Bellido,$^{23}$
D.~Gruen,$^{12}$
R.~A.~Gruendl,$^{24,25}$
G.~Gutierrez,$^{3}$
S.~R.~Hinton,$^{26}$
D.~L.~Hollowood,$^{27}$
K.~Honscheid,$^{28,29}$
D.~J.~James,$^{30}$
K.~Kuehn,$^{31,32}$
C.~Lidman,$^{33,34}$
M.~Lima,$^{35,10}$
J.~L.~Marshall,$^{36}$
J. Mena-Fern{\'a}ndez,$^{37}$
R.~Miquel,$^{38,16}$
J.~Myles,$^{39}$
R.~L.~C.~Ogando,$^{40}$
A.~Pieres,$^{10,40}$
A.~A.~Plazas~Malag\'on,$^{41,42}$
A.~K.~Romer,$^{43}$
E.~Sanchez,$^{20}$
D.~Sanchez Cid,$^{20}$
B.~Santiago,$^{44,10}$
M.~Schubnell,$^{1}$
I.~Sevilla-Noarbe,$^{20}$
M.~Smith,$^{45}$
E.~Suchyta,$^{46}$
M.~E.~C.~Swanson,$^{24}$
N.~Weaverdyck,$^{1,47}$
P.~Wiseman,$^{45}$
and M.~Yamamoto$^{48}$
\begin{center} (DES Collaboration) \end{center}
}\vspace{0.3cm} 
\\
{\small\emph{(Affiliations are listed at the end of paper)} }
}

\pubyear{2024}

\begin{document}

\label{firstpage}
\pagerange{\pageref{firstpage}--\pageref{lastpage}}
\maketitle

\begin{abstract}
    Cosmological analyses using galaxy clusters in optical/NIR photometric surveys require robust characterization of their galaxy content.
    Precisely determining which galaxies belong to a cluster is crucial. 
    In this paper, we present the \textbf{CO}lor \textbf{P}robabilistic \textbf{A}ssignment of \textbf{C}lusters \textbf{A}nd \textbf{BA}yesia\textbf{N} \textbf{A}nalysis (Copacabana) algorithm. 
    Copacabana computes membership probabilities for {\it all} galaxies within an aperture centred on the cluster using photometric redshifts, colours, and projected radial probability density functions. 
    We use simulations to validate Copacabana and we show that it achieves up to 89\% membership accuracy with a mild dependency on photometric redshift uncertainties and choice of aperture size. We find that the precision of the photometric redshifts has the largest impact on the determination of the membership probabilities followed by the choice of the cluster aperture size. We also quantify how much these uncertainties in the membership probabilities affect the stellar mass--cluster mass scaling relation, a relation that directly impacts cosmology. Using the sum of the stellar masses weighted by membership probabilities (\mustar{}) as the observable, we find that Copacabana can reach an accuracy of 0.06 dex in the measurement of the scaling relation. These results indicate the potential of Copacabana and \mustar{} to be used in cosmological analyses of optically selected clusters in the future.
\end{abstract}

\begin{keywords}
galaxies: clusters: general, methods: data analysis 
\end{keywords}





\section{Introduction}

Galaxy clusters have long been considered a promising astrophysical probes of dark energy  \citep{Albrecht2006, Allen2011, Dodelson2016} as their abundance as a function of redshift and mass is sensitive to the growth rate of structures in the Universe. Clusters are complementary to geometry based probes such as type-Ia supernovae and can be used to test different dark energy models \citep{Huterer2023}. The challenge in realizing this promise is to obtain a well-understood sample of clusters with unbiased mass measurements of a few per cent precision across a wide range of redshifts ($z \sim 0-1$) and masses ($M \sim 10^{13}-10^{15} \Mo$). In galaxy clusters, more than 80\% of the mass is in the form of dark matter while 5\%--15\% is diffuse hot gas, and only 1\% -- 5\% is in galaxies \citep{Gonzalez2007,Gonzalez2013,Lagana2013,2017ApJ...842...88S,2019SSRv..215...25P,2020A&ARv..28....7U}. Apart from a few extremely massive clusters for which direct total mass measurements via gravitational lensing are possible, we rely on indirect scaling relations to infer cluster masses. \je{For instance, the hot intracluster gas has two main observational signatures: a thermal bremsstrahlung (X-ray) emission and the Sunyaev-Zeldovich (SZ) effect. Although these signals correlate strongly with cluster mass, they are reliably detectable only for clusters at the high mass end  where the gas temperature and density are highest \citep[e.g.][]{Sarazin1988, Bleem2020, Klein2022}. } 
 
 Large optical imaging surveys such as the Sloan Digital Sky Survey \citep[SDSS;][]{2000AJ....120.1579Y} and the Dark Energy Survey \citep[DES;][]{2005astro.ph.10346T} \je{have produced} samples of tens of thousands of clusters with masses \je{$\lesssim 10^{14} \Mo$} \citep{Rykoff2014SDSS, Rykoff2016}.  For these low-mass galaxy clusters, the galaxy content is crucial to unlocking their potential for cosmology \citep{Wu2021}. Establishing observable quantities that correlate with cluster masses is a challenge. One such quantity is richness ($\lambda$), defined as the \je{probability-weighted sum} of \je{red-sequence galaxies} identified and selected by cluster finding algorithms such as redMaPPer \citep{Rykoff2016}. Richness is an empirical mass proxy optimized to find clusters and \je{is  correlated with the cluster mass} \citep{Rozo2009a,Rykoff2012}. This quantity relies on a linear colour-magnitude relation known as the red sequence. 

\je{The formation and evolution of the red-sequence are not well understood \citep{Butcher1984, Andreon2006, DeLucia2007, Cooper2007, Puddu2021}. This lack of understanding poses a challenge to optical and NIR wavelength cluster cosmology programs that use red-galaxy counts as the mass-proxy as systematic uncertainties associated with the mass-proxy scaling relations dominate the error budget \citep{2022arXiv220305416W}. One promising avenue to address this challenge is the development of a mass-proxy that includes all of the galaxy content of the clusters. Such a mass-proxy has a stronger theoretical foundation and, thus, can be simulated and studied more easily than the red sequence \citep[e.g.:][]{Anbajagane2020}.} 


\je{Mass measurements can be biased at the low richness end if selection effects are not treated properly. Significant effort has been made to understand the systematics uncertainties associated with richness. Projection effects \citep{2019MNRAS.482..490C, 2021MNRAS.505...33M} and optical selection bias are the leading terms contributing to the cluster mass systematic uncertainties \citep{2020MNRAS.496.4468S, 2022arXiv220305416W}. }

\je{Recently, alternative mass proxies have been studied, in particular, those derived from the stellar mass \citep[e.g.][]{Andreon2012, Pereira2018, Bradshaw2019arXiv} and intracluster light \citep[e.g.][]{2022MNRAS.515.4722H, Golden-Marx2023}.} In \citet{Pereira2018, Palmese2019}, we designed a novel mass proxy around the stellar mass content of the galaxies in a cluster. Defined as the weighted sum of the stellar masses, \mustar{} uses a more complete representation of the full population of \je{the cluster galaxies} than red-galaxy count methods.  
\je{In \cite{Palmese2019}, we used an X-ray sample of clusters to compare \mustar{}--mass observable and the $\lambda$ method, finding that both  present a similar scatter in total mass (see their Figure 6). In \cite{Pereira2020}, we performed a detailed weak-lensing mass calibration of \mustar{} and we found that the precision of the mass--\mustar{}--$z$ scaling relation was comparable to the one obtained by \cite{2019MNRAS.482.1352M} for a mass--richness--$z$ relation. Those results indicate that \mustar{} has the potential to become a competitive mass proxy for cluster cosmology.}

In this study, we introduce a new methodology called \textbf{CO}lor \textbf{P}robabilistic \textbf{A}ssignment of \textbf{C}lusters \textbf{A}nd \textbf{BA}yesia\textbf{N} \textbf{A}nalysis (Copacabana). This method assigns probabilities for all galaxies in the cluster region, independent of the \je{cluster finder selection}. Copacabana continues to improve the methodology of previous \mustar{}-based papers \citep{Pereira2018, Palmese2019}. 

Copacabana's membership assignment enables value-added information for cluster finders to be produced, even for those that do not rely on galaxy catalogues, such as SZ and X-ray. Moreover, Copacabana can be used to study the evolution and properties of galaxy clusters, such as their mass content and galaxy population. This algorithm is particularly useful for 
 X-ray and SZ-selected samples, as their stellar-mass function and the baryon content of the Universe, can then be analyzed \citep[][]{Gonzalez2013, Leauthaud2012, Kravtsov2018}.
 
In this work, we apply Copacabana to 
improve and validate the estimates of \mustar{}. We use the Buzzard DES Year 3 (Y3) simulations \citep{2022PhRvD.105l3520D} to validate our algorithm by quantifying the probability's impact on the scaling relation. In addition, we study the impact of  \je{the uncertainties in photometric redshifts} of three large photometric surveys: SDSS, DES, and the Legacy Survey of Space and Time \citep[LSST;][]{LSSTScienceBook}.

This paper is organized as follows: \autoref{sec:formalism} details the membership assignment methodology based on \je{the photometric redshift, colour and projected radial  probability distributions}; \autoref{sec:setup} gives an overview of the simulated dataset employed in our study; the validation of our algorithm is shown in \autoref{sec:results}, and conclusions are presented in \autoref{sec:conclusions}. Throughout this manuscript, we use logarithm in base ten ($\log$) and we adopt the cosmological parameter values: $\Omega_m = 0.3$, $\Omega_\Lambda = 0.7$, and $h=0.7$. 
\section{Formalism} \label{sec:formalism}
In this section, we outline the method used in the Copacabana algorithm\footnote{\url{https://github.com/estevesjh/ccopa}} for assigning membership probabilities to galaxies for a given cluster field. The main motivation  is to produce stellar mass estimations for cluster galaxies that only have photometric information. The method presented here improves and extends the algorithm used in \cite{Palmese2016, Pereira2018, Pereira2020}. This work has been inspired by previous papers by
\citet{George2011, Rykoff2014SDSS, Castignani2016}. Copacabana has two main differences relative to those papers: \je{colour distribution and optimisation of the cluster aperture}. The red and blue galaxy populations are modeled simultaneously. For the cluster aperture, we estimate \je{\Rcritical{}} which is defined as the radius containing 200 times the critical the density of the universe (at the cluster redshift). \Rcritical{} is inferred from the galaxy distribution around each cluster. 

\subsection{Membership Probabilities}\label{sec:membership}
For a given \je{cluster photometric field}, the galaxies present belong to only two classes: gravitationally bound systems and field galaxies, i.e. in the background and foreground galaxies. \je{For the bound systems}, we define member galaxies as the population inside the \Rcritical{} defined by host halo mass. We make this distinction given that there are projected correlated structures, filaments, infalling groups, and galaxies. In simulations we know which are the correlated galaxies, in data we do not.

We adopt a Bayesian inference approach to estimate membership probabilities. Within this framework, the probability of a galaxy being a member of the cluster is in general:

\begin{equation}
{\rm
    P( member|data) = \frac{ P(data|member) P(member)} {P(data)}} \; , 
\label{eq:member1}
\end{equation}

where ``data'' represents the galaxy input variables: cluster-centric distance, photo-z and colour. The term $\rm P(data|member)$ is our likelihood distribution, which is modeled as the product of the distributions of our input variables (described in detail in \autoref{subsec:cluster_likelihood}). The prior $\rm P(member)$ is defined as the ratio of the number of member galaxies and the total number of galaxies, i.e., 
 $\rm n_{C}/(n_{C}+n_{F})$,  where $\rm C$ and $\rm F$ indicate cluster and field, respectively. 
 
 The denominator, $\rm P(data)$, is the probability of the union $\rm (C \cup F)$ of the two groups:
\begin{equation}\label{eq:member2}
\rm
    P( data) = P(data|member) P(member)+P(data|field) P(field),          
\end{equation}
where  $\rm p(data|field)$ is the observed field distribution and $\rm P(field)$ is  $\rm 1-P(member)$.

\subsubsection{Cluster Likelihood}\label{subsec:cluster_likelihood}
%
The likelihood depends \je{on the joint} radial, photometric redshift and colour distributions of the cluster galaxies:

\begin{equation}
\rm
    P(data|member) = P(R,z_p,c|member) \; ,
\end{equation}

where the term on the right side is given by the product of each variable (radius R, photometric redshift z$_{\mathrm p}$ and colour c):  
$$\rm P(R,z_p,c|member) = P(R|member) P(z_p|member) P(c|z_p, member)  $$

{We assume that these variables are independent. 
This assumption is an approximation. A potential improvement, not explored in this paper, would be to model the joint color and radial probabilities since galaxies are redder at the cluster centre.} 
Combining \autoref{eq:member1} and \autoref{eq:member2} we find:

\begin{equation}\label{eq:memb_prob}
\rm 
    P(member|R,z_p,c) = \frac{P(R, z_p, c|member) \times P(member)}{Q},
\end{equation}
where Q comes from the law of total probability:  

$$\mathrm{Q = P(member)P(R, z_p, c|member) + P(field) P(R, z_p, c|field)}.$$ 
In principle, the full membership probability with the three variables has more potential constraining power than probabilities using fewer variables. For some cases,  though, it might be useful to separate the impact of the colour and photo-z variables. For example, we could study the impact of photo-z outliers in cluster galaxies or for blue BCGs. For this reason, we also compute the probability for each model variable: 

\begin{equation}\rm
    P(member|R) =  \frac{P(R|member) \times P(member)}{
                    P(member)P(R|member) + P(R|field) P(field)} \;,
\end{equation}
\begin{equation}  \rm     
    P(member|z_p) =  \frac{P(z_p|member) \times P(member)}{
                    P(member)P(z_p|member) + P(field)P(z_p|field)} \;, 
\end{equation}
\begin{equation}  \rm     
    P(member|c) =  \frac{P(c|member) \times P(member)}{
                    P(member)P(c|member) + P(field)P(c|field)} \;. 
\end{equation}

In this formalism, the probabilities 
allow flexibility for the user to drop the 
selection in a given variable
if needed. As an example, in the case of group galaxies, the assumption of a Navarro-Frenk-White (NFW) profile might not be applicable, and we might, therefore, remove the probability related to the radius R, so removing the radial filter. In the next section, we present the definition of each probability.
%
\subsubsection{Radial Filter}
%
We assume that the cluster galaxy radial distribution is a projected NFW profile \citep{Wright2000GRAVITATIONALHALOS}, and  a constant radial distribution for the background. In our case, the NFW profile density  has the form:

\begin{equation}\label{eq:NFW}
 \Sigma (R) = 
 \begin{cases}
     \frac{2\rho_s R_s}{r^2-1} \bigg[ 1 - \frac{2}{ \sqrt{r^2-1} } \arctan{\sqrt{\frac{r-1}{r+1}}} \bigg] & r>1 \\
     \frac{2\rho_s R_s}{3}                                                                                & r=1 \\
     \frac{2\rho_s R_s}{r^2-1} \bigg[1-\frac{2}{\sqrt{1-r^2}}\arctanh{\sqrt{\frac{1-r}{r+1}}} \bigg]      & r>1 \; , \\
 \end{cases} 
\end{equation}

where $r=R/R_s$ is the dimensionless radial distance, $\rho_s$ is the density scale parameter, and  $R_s = R_{200c}/c_{200}$, where $c_{200}$ is the concentration parameter, is a characteristic radius. 
To convert the surface mass density profile to a radial probability density function (PDF), we compute the normalization factor

\begin{equation}
\rm
Norm ={\int_0^{R_{200}}{2 \pi R' \Sigma (R')} dR'} \; 
\end{equation}
such that 

\begin{equation}
    P(R|\rm{member}) = \Sigma(R,R_{200},c_{200})/ \rm{Norm} \; .
\end{equation}

The NFW has two free parameters, the radius $R_{200}$ and the concentration $c_{200}$. We infer $R_{200}$ using a halo occupation distribution (HOD) model (see \autoref{sec:HOD_model}) and we set $c_{200}=3.59$ as this was shown to be a good fit for halos in this mass range selected in the DES Science Verification dataset \citep{Hennig2017}. \je{A common assumption is a surface constant background density $\Sigma_{\rm field}(R) = n_{bkg}$ \citep{Rykoff2016}. As a result, the field radial density probability, $\rm P(R| field)$, is   a constant value determined by the normalization $\rm \int_{0}^{R_{max}} 2 \pi R^\prime \Sigma_{\rm field} dR^\prime = 1$.} 

\subsubsection{Photometric Redshift Distribution}
If the galaxy is assumed to be in the cluster we can set its probability density function to be at the cluster redshift. In the scenario where all member's redshifts are known, the cluster galaxy redshift distribution can be described by a normal distribution, with mean $z_{\rm cls}$ and standard deviation $\sigma_{\rm{cls}}$. In the context of photometric redshifts for clusters, the cluster redshift uncertainty $\delta z$ is much larger than $\sigma_{\rm{cls}}$. For instance, the typical redMaPPer cluster error is $\delta z = 0.01 (1+z_{\rm cls})$ \citep{Rykoff2016}, as opposed to a spectroscopically derived $\sigma_{\rm{cls}}\approx 0.001$. In this regime, the galaxy photo-z distribution can be described analogously by: 

\begin{equation}\label{eq:prob_redshfift} 
    P\left(z_p|\text {member}\right)=\frac{1}{\sqrt{2 \pi \delta z^2}} \mathrm{e}^{-\left(\mathrm{z}_{\mathrm{cls}}-\mathrm{z}_{\mathrm{p}}\right)^2 / 2 \delta z^2}
\end{equation}

where $z_p$ is the galaxy photo-z. Despite the simplicity of the model, it is a robust estimator, as shown by \cite{Castignani2016}. 
\je{For a given photo-z sample, the galaxy population photo-z can be biased relative to the cluster redshift. In such a case an offset to the cluster redshift can be applied \citep{Aguena2021}.}

If the galaxy is assumed to be in the field,
we need the field photometric redshift distribution. We use Gaussian Kernel Density Estimation (KDE) to estimate the photometric redshift density distribution in a ring around the cluster centre: 

\begin{equation}
    P(z_p|\mathrm{field}) = {\mathrm{KDE} }(z_p|R,h).
\end{equation}

The KDE has a bandwidth $h$ (i.e. the width of the Gaussian kernel) as a free parameter. The choice of a proper bandwidth depends on the shape of the underlying distribution and the number of objects available to construct the estimator. Assuming a normal distribution, we take the optimal bandwidth as:

\begin{equation}\label{eq:bandwidth}
    h = \left(\frac{4 \hat{\sigma}}{3n} \right)^{1/5}
\end{equation}

where $\hat{\sigma}$ is the sample standard deviation and $n$ denotes the number of objects. This ``Scott’s rule'' bandwidth is in common usage in statistics \citep{Scott1992}. 

\subsubsection{Color Distribution}
Generalizing from \citet{Pereira2020}, we add  color probabilities by using a 
color distribution subtraction method. The cluster color distribution is computed by subtracting the background color distribution from the total color distribution using a KDE: 

\begin{equation}
\begin{aligned}
N_{\rm cls} P(c|\rm{member}) ={} & N_{\rm total}{\rm KDE}(c|h_{\rm eff},\rm{total})\\
                  & - N_{\rm bkg}^{\prime}{\rm KDE}(c|h,\rm{field}).
\end{aligned}    
\end{equation}

The KDEs are weighted by the photo-z probability weights and $h_{\rm eff}$ is a Scott rule bandwidth divided by 10, a factor that is arbitrarily chosen to avoid over-smoothing. We only take the excess in the subtraction, so there are no negative values. 
For  $h_{\rm eff}$ the number of objects is computed in terms of an effective number 

$$n_{\rm eff} = \frac{(\sum{\rm weights} )^2}{\sum{\rm weights^2}}.$$ 

After the subtraction, we normalize $p(c|z, R)$ to unity. We choose one color filter at a time to be our color model: 

\begin{equation}
    c(z) = 
    \begin{cases}
    (g-i) \quad \text{for }z \leq 0.35\\
    (r-z) \quad \text{for }z > 0.35.
    \end{cases}
\end{equation}

The color filter changes at $z=0.35$ due to the $4000$ \r{A} break exiting the red edge of the $g$ band. In comparison with \citet{Pereira2020}, the addition of color probabilities in general improved the performance. 

\subsection{Number Densities}

Computing probability weighted number density is the first step in our algorithm, and is \je{used in all the steps were galaxies are counted, i.e. the background subtraction, the estimation of the radius $R_{200}^{\rm HOD}$ and the color model.}    
\je{The first step is to count galaxies around the cluster center, within a cylinder of radius $R_{200}$ and height $2\sigma_{z}(1+z_{\rm cls})$, where $z_{\rm cls}$  is the redshift of the cluster.  The galaxies are counted using a probabilistic weight.} The probability of a galaxy being at the cluster redshift, $P_{z_{0}}$, \je{is the integral of the galaxy's photometric redshift distribution ($\Pi(z)$) around the cluster redshift}. The limits of integration are chosen in a window of $2\sigma_{z} (1+z_{\rm cls})$ around the mean value: 

\begin{equation}
    P_{z_{0}} = \int_{z_{-}}^{z_{+}} \Pi (z) \; dz, \qquad z_{\pm}=z_{\rm cls} \pm 2\times \sigma_{z,0} (1+z_{\rm cls}) \; ,
\end{equation}
The limits of the integration depend on the photo-z precision for a given redshift, $\sigma_{z,0}(z)$. This counting method selects galaxies near the cluster redshift while avoiding a sharp, arbitrary cutoff in redshift space. 

This probability weight is the first step in our algorithm and is \je{used in all the steps where galaxies are counted, i.e. the background subtraction, the estimation of the radius $R_{200}^{\rm HOD}$ and the color model.}    

\subsection{Background Subtraction}
Background subtraction is an essential step for computing membership probabilities. There are two methods traditionally used: global and local subtraction. The global background subtraction method  assumes that the background density only depends on the redshift \citep[e.g.][]{Rykoff2014SDSS}. However, this assumption must be invalid as  clusters are nodes of the cosmic web. Consequently, galaxies in the line of sight are more likely to be assigned as members. %

In our work, we choose to use a local background subtraction method that probes the surroundings of each cluster region. \je{Specifically, we compute the galaxy density in an annulus centered on the cluster with inner and outer radii of $4$ Mpc and $6$ Mpc, respectively. The inner radii are always larger than \Rcritical{}, even for the most massive clusters for which \Rcritical{} is approximately $3$ Mpc. Although scaling the radii with the cluster radius would be an optimal choice, we prefer to use fixed values since our \Rcritical{} estimation depends on the background density. }



\subsection{\Rcritical\ estimator: HOD Model}\label{sec:HOD_model}

Clusters don't have obvious edges, and various investigations define cluster apertures differently.
For instance, redMaPPer \citep{Rykoff2014SDSS} assumes an aperture that scales with richness $\lambda$, and AMICO \citep{2018MNRAS.473.5221B} assumes a fixed aperture \je{corresponding to} a cluster with $M_{\rm 200,c}=10^{13.5} \; M_{\odot}$. In this work, we introduce a new cluster aperture estimator based on a Halo Occupation Distribution (HOD) model, which is independent of our mass proxy. 


Our aperture estimator uses the galaxy number density profile of a HOD model. A given HOD model provides the number of halo galaxies as a function of mass which allows us to convert the number density profile to a mass density profile. Assuming spherical symmetry, we can calculate the mass density, $\rho$. We can make a rough estimation of $R_{200}$ by interpolating the mass density profile as a function of radii. By definition, where the mass density profile is 200 times the critical density, we have our aperture estimation, $R_{\rm 200c}^{\rm HOD}$.
For this work, we adopt the HOD model of \cite{Tinker2012}. The model consists of a relation between the number of central ($N_{\rm cen}$) and satellite galaxies ($N_{\rm sat}$) inside a halo of given mass ($M_{\rm 200,c}$) and below a given luminosity threshold. The occupation function for central galaxies takes the form: 

\begin{equation}\label{eq:hod_ncen}
\langle N_{\rm cen} \rangle_M = \frac{1}{2} \Bigg[ 1 + \erf{ \Bigg( \frac{ \log M - \log M_{\rm min}}{\sigma_{\log M}} \Bigg) } \Bigg],
\end{equation}

where $M$ is the halo mass, $M_{\rm min}$ represents the halo mass at which the probability of containing a central galaxy is $50$ per cent, and $\sigma_{\log M}$ accounts for the scatter in halo mass at a fixed luminosity of the galaxy population. The occupation function for satellites is given by a power law: 

\begin{equation}\label{eq:hod_nsat}
\langle N_{\rm sat} \rangle_M = \langle N_{\rm cen} \rangle_M \times \Bigg(\frac{M}{M_{\rm sat}}\Bigg)^{\alpha_{\rm sat}} \exp{\Bigg(-\frac{M_{\rm cut}}{M}\Bigg)},
\end{equation}

where $\alpha_{\rm sat}$ is the slope at high halo masses, with an exponential cutoff at halo masses below $M_{\rm cut}$, and $M_{\rm sat}$ is the characteristic halo mass for satellites. Combining the central and satellite occupation functions produces a total occupation function of the form 

\begin{equation}
\langle N_{\rm tot} \rangle_M = \langle N_{\rm cen} \rangle_M \times \Bigg[1 + \Bigg(\frac{M}{M_{\rm sat}}\Bigg)^{\alpha_{\rm sat}} \exp{ \Bigg( -\frac{M_{\rm cut}}{M} \Bigg) }\Bigg].
\end{equation}

There are five free parameters this  model. 
The best-fit values for these parameters (given in Table 4 of \cite{Tinker2012}) were derived from the SDSS dataset, using the maxBCG cluster sample \citep{2007ApJ...660..239K}, \je{with an absolute magnitude cut $M_r^{0.1} \leq -19.5$. We use these values as a reference since they are close to the $0.2 L_\star$ cut applied here \citep{Rykoff2012}. }



We found that this approach results in  \Rcritical{} estimates that are biased low. To account for this bias, we introduce a calibration factor which can be computed as the mean ratio of our predictions and the actual \Rcritical{} values in the simulations: 

\begin{equation}\label{eq:hod_correction_factor}
    \rm 
    \eta_{HOD} = \frac{R_{HOD,200c}}{R_{200c,true}} \qquad \text{where}, \qquad R^3_{HOD, 200c} = 200 \frac{M(N_{200c})}{4\pi \rho_c /3} \; .
\end{equation}

This calibration factor is independent of redshift, as we wish to use only one factor for the entire population. 

\subsection{Stellar Mass Estimation: BMA}\label{sec:bma-stellar-mass}

The code we call  BMA\footnote{\url{https://github.com/apalmese/BMAStellarMasses}} is a Bayesian model averaging code \citep[see e.g.][]{Taylor2011} applied to the output of a stellar population synthesis code, and developed into a pipeline \citep{Palmese2019}. We use the stellar population synthesis code FSPS \citep{Conroy2009, Conroy2010} to evaluate a 5-dimensional space of quantities, resulting in 24 models. We choose the models evaluated at the cluster redshift for a given galaxy with apparent magnitude, colours,
and photo-z. Then the likelihood of each model given the galaxy
magnitudes, colours, and errors is computed. The properties of interest from the models, e.g. stellar mass, are then
computed as the likelihood-weighted sum over all models, a Bayesian model average. The code was validated on the Millennium simulations \citep{2005Natur.435..629S} and on the COSMOS dataset \citep{Laigle2016}. See \cite{Palmese2019} for a full description of the BMA methodology.

\subsection{\ensuremath{\mu}$_{{\ensuremath{\star}}}$ Estimator}
A key motivation for the development of Copacabana is to improve measurements of the mass proxy \mustar. To accomplish that, we focus on the membership probabilities. \mustar{} is defined in a probabilistic manner: 

\begin{equation}\label{eq:mu-star}
    \mu_{\star} = \sum_{i} P_{{\rm mem},i} M_{\star,i} \; \; \text{for} \; \; R \leq R_{\rm aper} \; ,
\end{equation}
where $M_{\star}$ is the galaxy's stellar mass, $ P_{\rm mem}$ is the membership probability (\autoref{eq:memb_prob}). \je{$R_{\rm aper}$ is the cluster radius aperture which is \Rcritical{} if not defined otherwise}. A galaxy's stellar mass is estimated from photometric data assuming it is at the cluster redshift via BMA.

\section{Validation Setup}\label{sec:setup}

\subsection{Data}
\je{To validate Copacabana,} we use the Buzzard v2.0 simulations \citep{DeRose2019arXiv}, meant to correspond with the DES Y3 area. The dataset consists of synthetic dark matter simulations with galaxy information added by the AddGals algorithm \citep{Wechsler2022}. This procedure places galaxies onto the dark-matter-only simulation, weighted by local dark matter density, matching the observed luminosity function and luminosity-dependent two-point correlation function.

For context, we briefly describe how the galaxies are pasted onto the dark matter particles. \je{First, AddGals creates} a catalogue of galaxies based on the luminosity function, $\phi(M_r)$, performing subhalo abundance matching between a small high-resolution N-body simulation and the observed SDSS luminosity function in the r-band. The algorithm calibrates a relation (not a HOD) for the central and \je{the non-central galaxies}, which is then evolved using a functional form. The model of the central galaxies is a log-normal distribution at a fixed halo mass and redshift, $P(M_{r,{\rm cen}}| M_{\rm vir}, z)$, and the model of the non-centrals is based on the local density. These relations are then used to assign galaxies to resolved haloes or dark matter particles in a large light-cone simulation with a lower resolution. This modelling scheme was chosen such that it predicts the clustering in SDSS to high precision \citep[e.g.][]{Conroy2006, Reddick2013, Lehmann2017}. \je{Second, colours are assigned using a spectral energy distribution (SED), chosen such that the simulation matches the SED distribution (at fixed luminosity and galaxy density) measured in the SDSS data.} 

\begin{figure}
    \centering
    \includegraphics[width=0.5\textwidth]{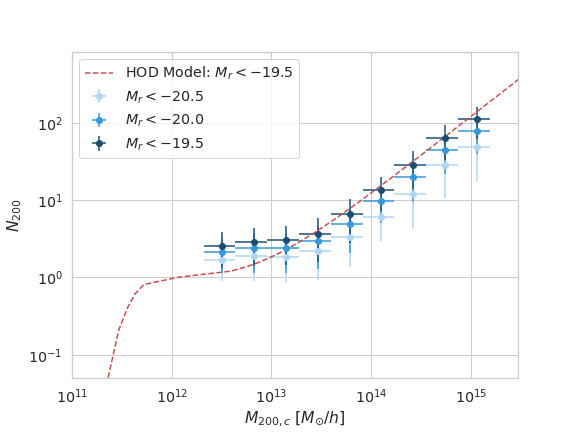}
    \caption{Number of galaxies inside $R_{\rm 200,c}$ ($N_{200}$) as a function of the halo mass $M_{\rm 200,c}$, for three absolute magnitude cuts in the r-band. The Buzzard galaxy distribution follows the \citet{Tinker2012} model closely (red dotted line) when both apply the same magnitude selection of $M_r^{0.1} \leq -19.5 \, \text{ mag}$ (dark blue points).}\label{fig:hod_plot}
\end{figure}

The galaxies in Buzzard, unlike many cosmological simulations, are not placed using a HOD prescription. Nonetheless, the halo occupation distribution on Buzzard follows the \citet{Tinker2012} model closely for an absolute magnitude selection of $M_r^{0.1} \leq -19.5$, as we can see in \autoref{fig:hod_plot}. In addition, \citet{Zacharegkas2022}, using a Buzzard redMaGic galaxy selection, found acceptable HOD model fits (\autoref{eq:hod_ncen}, \autoref{eq:hod_nsat}) to the galaxy distribution. 

\subsubsection{Sample Selection}
In order to accurately assess the performance of our code across the halo redshift and mass ranges, \je{we select} with bin-dependent uniform probability 
 \je{ halos in bins of redshift and logarithmic mass, i.e. [$\log M_{\rm 200,c}, z$].} The aim is to ensure we have the same number of halos for each bin, which we do not quite achieve in high mass bins due to a lack of clusters. This approach prevents our assessment from being biased by the low end of \je{the halo distribution.}

\begin{figure}
    \centering
    \includegraphics[width=0.5\textwidth]{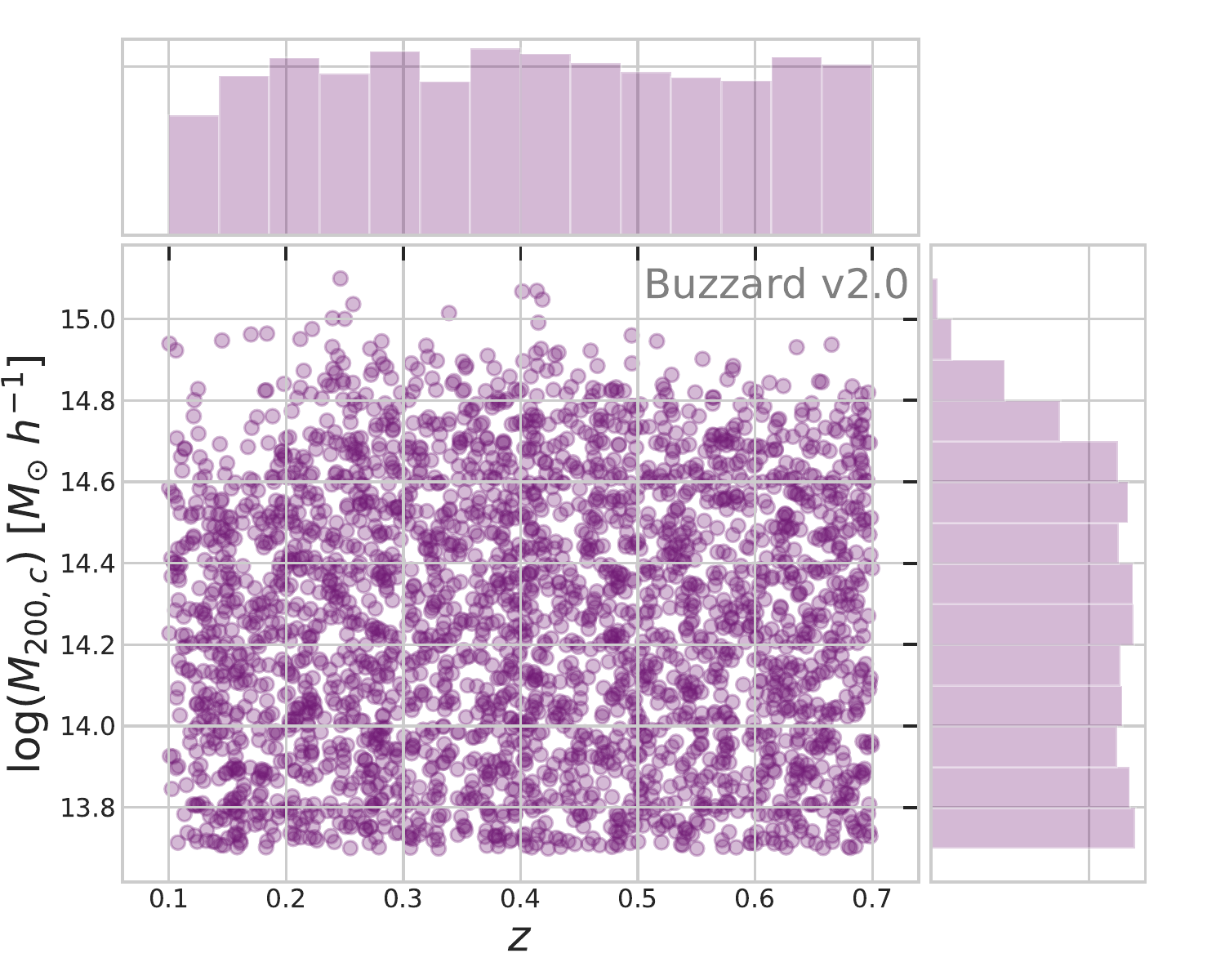}
    \caption{Uniform selection of 2,200 Buzzard v2.0 halos on a halo mass--redshift grid. The upper and right panels are the redshift and halo mass distributions, respectively.}
\label{fig:sample_selection}
\end{figure}

The data chosen by our selection is presented in \autoref{fig:sample_selection}, where 2,200 halos are plotted in $(z, \log M_{\rm 200,c})$ space with histograms on the $x$-- and $y$--axis. The limits of the sample are $z \in \left[ 0.1, 0.65 \right]$ and $(\log M_{\rm 200,c})>13.5 \; M_{\odot}/h$. The choice of the redshift range follows the DES cluster cosmology analysis \citep{2020PhRvD.102b3509A}. The halo mass threshold \je{is similar to that adopted by other}
cluster finder algorithms \citep[e.g.:][]{Castignani2016,Bellagamba2019}. 
Overall, there is a uniform selection, except for the highest mass bins, where there are not enough systems.

\subsubsection{Simulated Photo-z}\label{sec:photo-z-samples}
To validate the Copacabana algorithm with respect to photo-z, \je{we add offsets to the simulated galaxy redshifts. In detail, we draw a random offset following:
\begin{equation}
    z_p = \mathcal{N}\left(z, \sigma_{z,0}(1+z) \right) \; ,
\end{equation}
where $\sigma_{z,0}$ is the photo-z precision that correspondents to a typical photo-z  error.}

\je{We simulate three different levels of uncertainty: $\sigma_{z,0} = 0.01, 0.03, \text{ and } 0.05$ as an ideal, a realistic, and a pessimistic case, respectively.} These choices mimic three different surveys: LSST \citep{LSSTScienceBook}, DES \citep{Gschwend2018, Aguena2021} and SDSS \citep{Carliles2010}, respectively. 

In the context of clusters, the main differences between simulated Gaussian photo-z's with real data photo-z's are the bias and the presence of outliers. For instance, \cite{Aguena2021} using the WaZP cluster catalogue studied the differences of redMaPPer cluster redshift with the ones derived from the DNF photo-z algorithm in the DES Y1 \citep{Gschwend2018}. They quantified a redshift bias that is less than $0.003 \times (1+z) $. For future applications of Copacabana on data, a description of the bias between the photo-z sample employed and the cluster redshift must be taken into account as a bias on the cluster photo-z distribution (\autoref{eq:prob_redshfift}).

\subsection{Validation Metrics}
\subsubsection{Assessing \ensuremath{\mu}$_{{\ensuremath{\star}}}$ precision}
To validate our \mustar\ estimation, we compare it with the simulation cluster member stellar masses. For this purpose, we define \mustarTrue, 
\begin{equation}\label{eq:mu-star}
    \mu_{\star,\rm true} = \sum_{i \in {\rm members}} M_{\star,i} \; \; \text{with} \; \; R \leq R_{\rm aper} \; ,
\end{equation}
as the sum of the cluster members' stellar masses, where the ``true'' members are defined as the galaxies inside the three-dimensional \Rcritical{} distance from the cluster centre. \je{In other words, we don't consider the line-of-sight infall galaxies, nor the gravitational status of the galaxy.}

Next
\je{we can define the ratio $ x \equiv \mu_{\star}/\mu_{\star, \rm true}$. Since our data has a non-Gaussian tail at very low richnesses, $N_{\rm gal} < 10$, a robust metric is adopted, the scaled median absolute deviation (MAD)}:
\begin{equation}\label{eq:metric_rms}
    \sigma_{\rm MAD} (\log(x)) = 1.48 \times \text{Median}\left( \left| \log(x) -  \text{Median}\left({\log}(x)\right) \right| \right) \; .
\end{equation}
Note that if $\log(x)$ follows a normal distribution $\sigma_{\rm MAD}$ is equal to the standard deviation. 

It is important to stress that 
our assessment is primarily on our estimator due to membership probabilities and we do not evaluate uncertainty due to stellar mass estimates.

\subsubsection{Assessing the accuracy of  \ensuremath{R_{200c}}}
To validate our estimates of \Rcritical{}, we use the current value from the simulation. $R_{200c,\rm true}$ was retrieved from the Buzzard truth table. \je{Analogously to \mustar{}, we evaluate \autoref{eq:metric_rms} with x set to the ratio of true versus measured \Rcritical{}. }

\subsubsection{Completeness and Purity}

The membership probabilities play a role in thresholds that distinguish the classes in the framework of classifying members and non-members of a given galaxy cluster. 
To assess the performance, we use metrics commonly used in statistical classification problems, purity ($P$) and completeness ($C$).These metrics rely on true positive (TP), false positive (FP), and false negative (FN) predictions:

\begin{equation}\label{eq:Purity_Completeness}
P = \frac{TP}{TP+FP} \quad \text{and} \quad C=\frac{TP}{TP+FN} ; ,
\end{equation}

TP represents correct positive predictions, while FP and FN refer to incorrect positive and negative predictions, respectively. \je{Purity indicates the proportion of positives that are cluster members. Completeness measures the fraction of true members that were successfully identified among all the selected galaxies.}

The overall accuracy of a classifier can be evaluated by:
\begin{equation}
    \text{accuracy} = \frac{TP+TN}{TP+FP+TN+FN} \, ,
\end{equation}
where TN is the true negatives, i.e. correctly identified field galaxies. 

\subsubsection{\ensuremath{\mu}$_{{\ensuremath{\star}}}$-Cluster Mass Scaling Relation}\label{sec:metrics_mass_observable_relation}

For photometric surveys, one of the main requirements for tight constraints on cosmological parameters is a mass proxy that predicts the cluster mass with significant accuracy and is robust against systematic effects. Here, we assess the possible impact of the membership probabilities on deriving cosmological results using \je{the relation between the weighted stellar mass and the total cluster mass, which we will refer to as the \mustar{} - cluster mass scaling relation.}

In simulations, the \mustar{} - cluster mass scaling relation is accessible since the halo mass is known. The probability of a given halo of mass $M_{\rm 200,c}$ to have a $\mu_{\star}$ value is generally modeled by a Log-Normal relation with mean:
\begin{equation}\label{eq:linear_relation}
\left< \log (\mu_{\star}) | M_{\rm 200,c} \right> = \alpha + \beta \log(M_{\rm 200,c}/M_p) \; ,
\end{equation}
with an associated intrinsic error $\sigma$. Here $\alpha$ is the intercept, $\beta$ is the slope, $M_p = 10^{15.5} M_{\odot}$ is the pivot mass. The inference of the model parameters is made by employing a hierarchical Bayesian algorithm \citep[\texttt{linmix};][]{Kelly2007}. The \texttt{linmix} algorithm allows us to include the error on the $y$--dependent variable, in our case, \mustar{}. 

%
\je{In general the scaling relation evolves with redshift.} We model redshift evolution by fitting the observable--mass relation in different redshift bins. We follow the standard parametrization of the evolution as proportional to the power of the scale factor $a \equiv 1/(1+z)$ or the dimensionless Hubble parameter $E(z) \equiv H(z)/H_0$ \citep{Evrard2014}. 
 
\section{Results} \label{sec:results}

In this section, \je{we examine the performance} of Copacabana using the Buzzard simulation. 
\subsection{Uncertainty in \ensuremath{\mu}$_{{\ensuremath{\star}}}$ Estimations}

We run Copacabana on the Buzzard v2.0 simulation using \je{the three values of $\sigma_{z,0}$} presented in \autoref{sec:photo-z-samples}. We employ the photometric stellar masses computed by BMA at the cluster redshift (\autoref{sec:bma-stellar-mass}). The stellar mass \mustar{}\ is computed within an aperture  \Rcritical\ estimated using the HOD model presented in \autoref{sec:HOD_model}. The membership probabilities are expected to depend mainly on the photo-z uncertainty.

\begin{figure}
    \centering
    \includegraphics[width=0.5\textwidth]{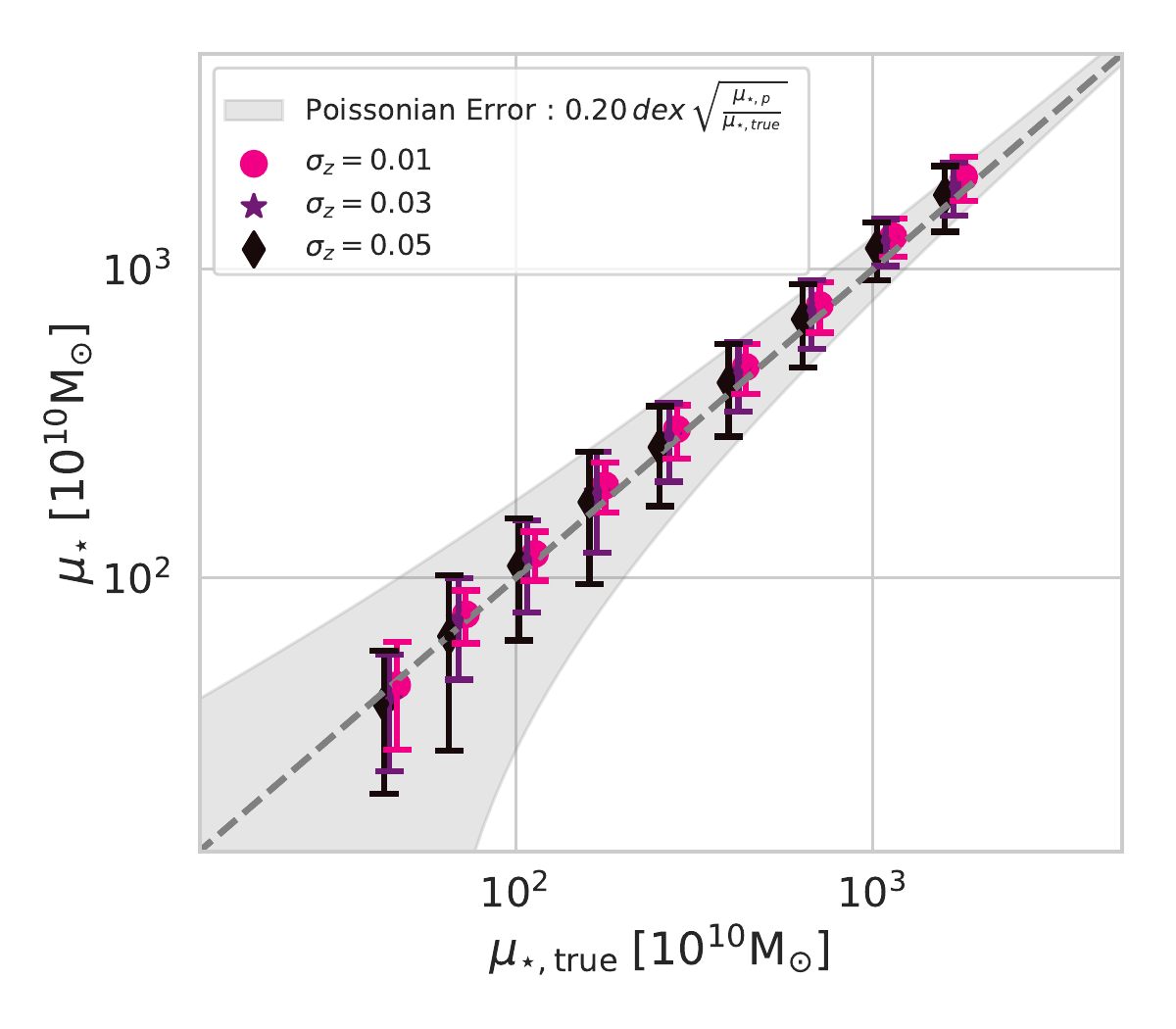}
    \caption{Copacabana predicted \mustar{} as a function of \mustarTrue{} (the sum of the stellar masses of the true cluster members within \Rcritical{}) for different values of $\sigma_{z,0}$, the uncertainty in the galaxy photometric redshifts. The binned points were slightly shifted for better visualization. The estimated values closely follow the one-to-one relation (grey dotted line). For comparison, we also show the range (in light grey) corresponding to a Poissonian error of $0.20 \text{ dex } \sqrt{\mu_{\star,p}/\mu_\star}$ where $\mu_{\rm \star,p}$ is the sample mean of $10^{12.22} M_\odot$. Note that our result is robust well below the threshold of $10^{12}M_\odot$ used in our previous work \protect\citep{Pereira2020}.  
    }\label{fig:scaling_relations}
\end{figure}

\autoref{fig:scaling_relations} shows the estimated \mustar{} versus \mustarTrue{} within \Rcritical{} for three values of $\sigma_{z,0}$. The \mustar{} values  for all halo mass regimes follow the \mustarTrue{} values closely. The \mustar{} errors are roughly within $0.20 \text{ dex } \sqrt{\mu_{\star,p}/\mu_\star}$, where  $\mu_{\rm \star,p} =  10^{12.22} M_\odot$ is the sample mean. 
Uncertainty in the position of the galaxies along the line-of-sight adds significant noise, especially for the low-mass halos as they have fewer galaxies. As a result, the quality of the photometric redshifts substantially impacts the \mustar{} measurements and particularly does at the low-mass end. Nevertheless, a sample with accurate and precise photometric redshifts can accurately predict \mustar{} down to the lowest bin, $5\times 10^{11} M_{\odot}$.

\subsubsection{Photometric Redshift Uncertainties And Cluster Apertures}\label{subsec:mu-star-precision-photoz-cluster-aperture}
In this section, we quantify \je{how the uncertainty in \mustar{} depends on $\sigma_{z,0}$ and the size of the apertures used to estimate \mustar{}}. We show that the photo-z quality is the main systematic on the \mustar{} error budget.

\begin{figure}
    \includegraphics[width=0.45\textwidth]{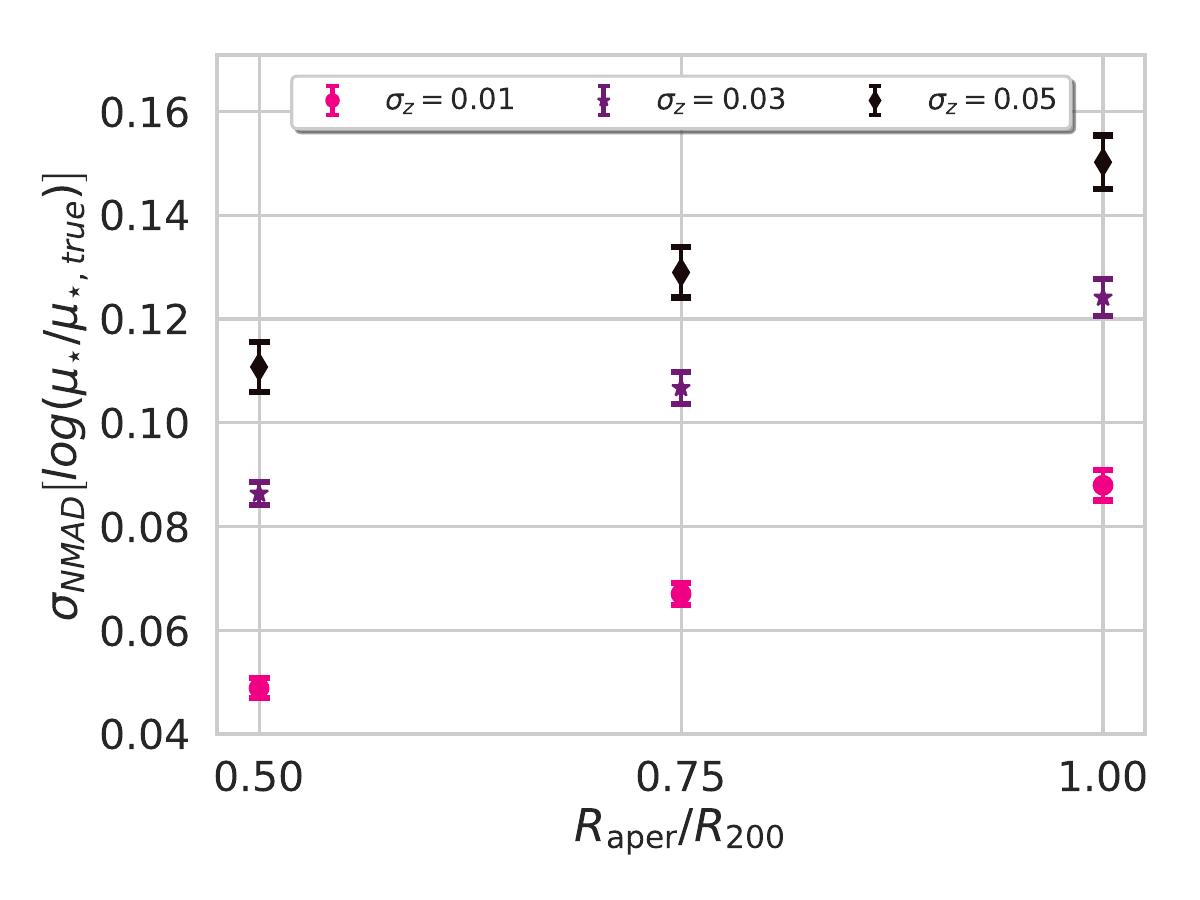}
    \caption{Copacabana $\sigma_{\rm MAD}$ error of the $\log$ \mustar$/$\mustarTrue{} as a function of the cluster aperture radius for three different photo-z precisions. The  $\sigma_{\rm MAD}$ error is always below $0.16$ dex, and it varies significantly with the photo-z quality (different markers). In particular, for an LSST-like photo-z (pink markers),  $\sigma_{\rm MAD}$ is $0.09$ dex. The precision can be improved considerably by defining a smaller cluster aperture since the density is higher in the cluster centre. The error bars were estimated by bootstrapping the cluster sample a thousand times.} \label{fig:raper}
\end{figure}

In \autoref{fig:raper}, we show the $\sigma_{\rm MAD}$ (defined in \autoref{eq:metric_rms}) as a function of the cluster aperture and photo-z uncertainty. \je{The accuracy of \mustar{} correlates linearly with the uncertainty in the photometric redshifts.} For instance, an improvement of a factor 5 (1.7) in $\sigma_{z,0}$ reduces the \mustar{} error to $0.15 \text{ dex}$ ($0.09 \text{ dex}$) when compared to $\sigma_{z,0}=0.05$. This improvement in the photometric redshifts has a big impact, especially for low-mass halos, see \autoref{fig:scaling_relations}. This result follows from the fact that the uncertainty along the line of sight is the major source of galaxy membership contamination. For example, a redshift error of $0.01\times (1+z)$ translates into a physical length of $\sim 40$ Mpc which is $\sim 10 - 50 \times R_{\rm 200,c}$.  

A second source of error is the cluster aperture, which has a non-negligible effect on the uncertainty in \mustar{}. As shown in \autoref{fig:raper}, smaller cluster apertures decrease the uncertainties in \mustar{}. This effect is driven by the higher galaxy density in the core. \citet{Lopes2020} showed that the local density is a very good indicator of membership galaxies in their well-characterized cluster sample.  


The choice of the cluster aperture is important for studying cluster cosmology. 
Ideally, we would expect that \mustar{} has the highest correlation with the halo mass when computed at \Rcritical{} \citep{Rykoff2012,Bradshaw2019arXiv, Song2020}. In \autoref{sec:mass_observable_relation} we discuss further the choice of the cluster aperture in terms of optimizing the scatter of the \mustar--$M_{\rm 200,c}$ relation. 

\subsubsection{Stellar Mass Estimation}
Thus far, the \mustar{}\ \je{uncertainty} was computed without taking into account the errors on stellar mass. The BMA stellar-mass error is around 0.2 dex \citep{Palmese2019}, validated using the stellar masses computed with multi-band data in 16 filters from UV to infrared of the COSMOS deep field \citep{Laigle2016}. The BMA \je{stellar-mass errors are comparable} to the uncertainties induced by the photo-z errors, see \autoref{fig:raper}. Therefore, they have a significant \je{impact}. To quantify this impact, \je{we add random noise normally distributed to the estimated stellar masses}. We assume the typical BMA error of 0.20 dex. This assumption should set \je{an upper bound on the uncertainty in \mustar{}. Because \mustar{}\ is dominated by high stellar mass (bright) galaxies that have lower mass uncertainties.}

The additional scatter on \mustar{} is $0.07 \pm 0.01 \, \rm{dex}$ for the three photo-z samples. This result is equivalent to adding the stellar mass error in quadrature, $\sigma_{\rm MAD}^2 + \sigma_{\rm BMA}^2$. The implication of this additional error for the DES-like photo-z case is that the \mustar{} error is at the same level as an SDSS-like photo-z. In future work, it would be important to reduce the uncertainty \je{on stellar masses to reduce the uncertainty on \mustar{}}.

\subsection{Precision of \Rcritical\ Estimations}\label{sec:r200_hod_results}

The new \Rcritical{} estimator is based on a HOD model, which is sensitive to the relation between the number of galaxies and the halo mass. To perform our measurements, we applied a calibration factor $\eta_{HOD}$ (\autoref{eq:hod_correction_factor}). \je{The derived correction factor is $ 0.63 \pm 0.11$ for the same HOD luminosity cut. We use all Buzzard halos with $M_{\rm 200,c}>5  \times 10^{13} M_\odot$ for this computation. Note that only this correction factor is set to calibrate the HOD relation.}

\begin{figure}
    \centering
    \includegraphics[width=0.5\textwidth]{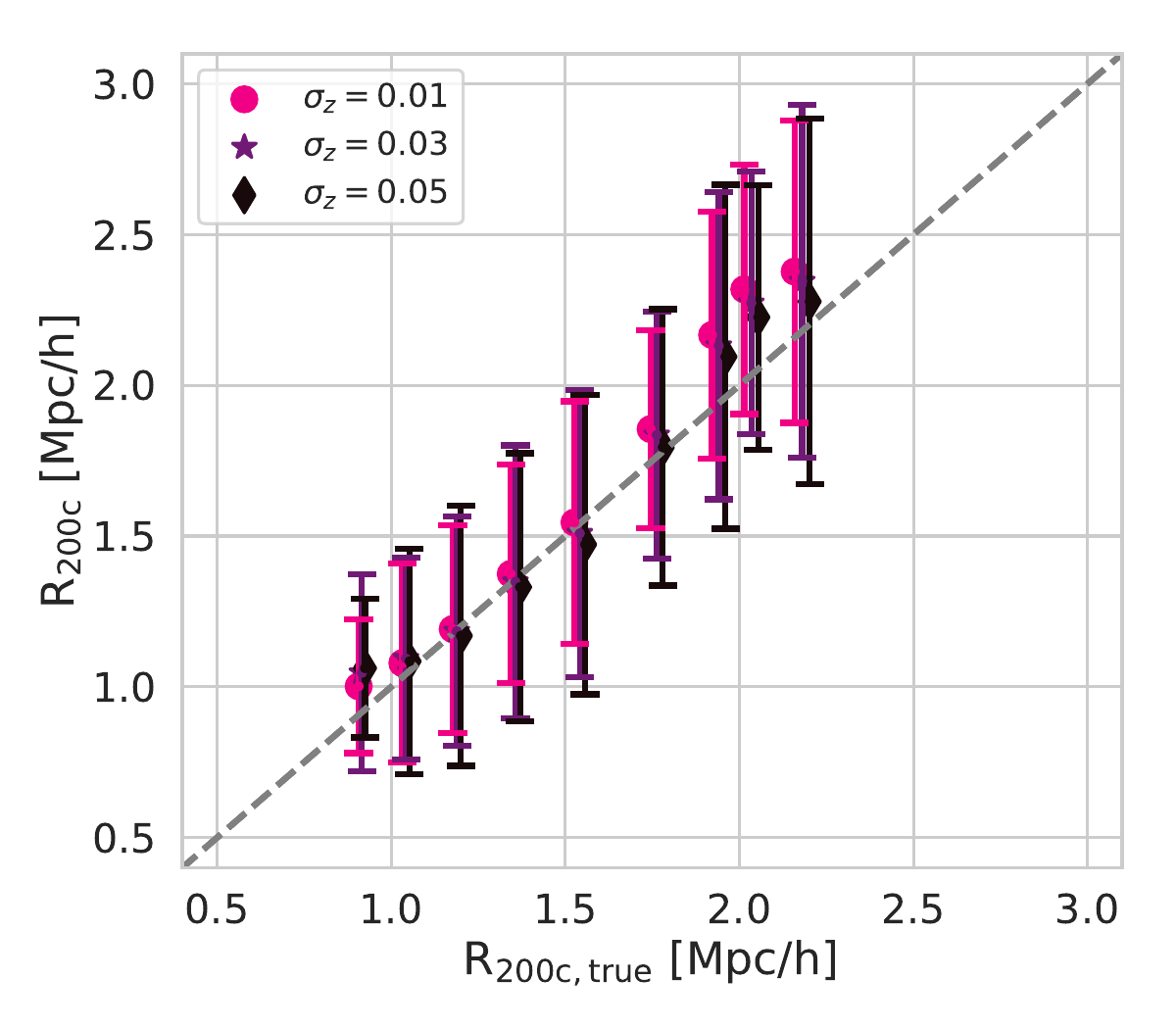}
    \caption{The Copacabana predicted \Rcritical{} based on a HOD model versus the true radius ($R_{\rm 200c, true}$) for three values $\sigma_{z,0}$.}\label{fig:rhod}
\end{figure}

Our estimator predicts the \Rcritical{} \je{for halo masses probed} in this study as shown in \autoref{fig:rhod} with a scatter around $30\%$ after calibration. Although the scatter is large, our estimated values correlate with the $R_{\rm 200c, true}$ for low and high-mass clusters, which makes them a good probe of the cluster size. Interestingly, the photo-z does not have a significant impact on the predicted values. Likewise, \je{the accuracy of the estimated radius is independent of cluster mass, unlike what we found for \mustar{}.} These results indicate that the scatter observed in \autoref{fig:rhod} comes mainly from another source, \je{likely the intrinsic error of the HOD relation. }


The significant uncertainties in our estimates of \Rcritical{} relate to the uncertainty of the HOD model itself. For instance, the HOD model halo mass scatter is $0.204 \; \rm dex$. Also, optical mass proxies, in general, have similar or higher intrinsic errors, e.g. the brightest cluster galaxy (BCG) stellar-mass proxy intrinsic scatter is $0.20$ \citep{Behroozi2010, Tinker2012}. In the future, if there are precise stellar mass measurements, the stellar-to-halo mass can be incorporated into our methodology \citep{Behroozi2010}. 


\subsection{Completeness and Purity of the Members List}
\begin{figure}
    \centering
    \includegraphics[width=0.5\textwidth]{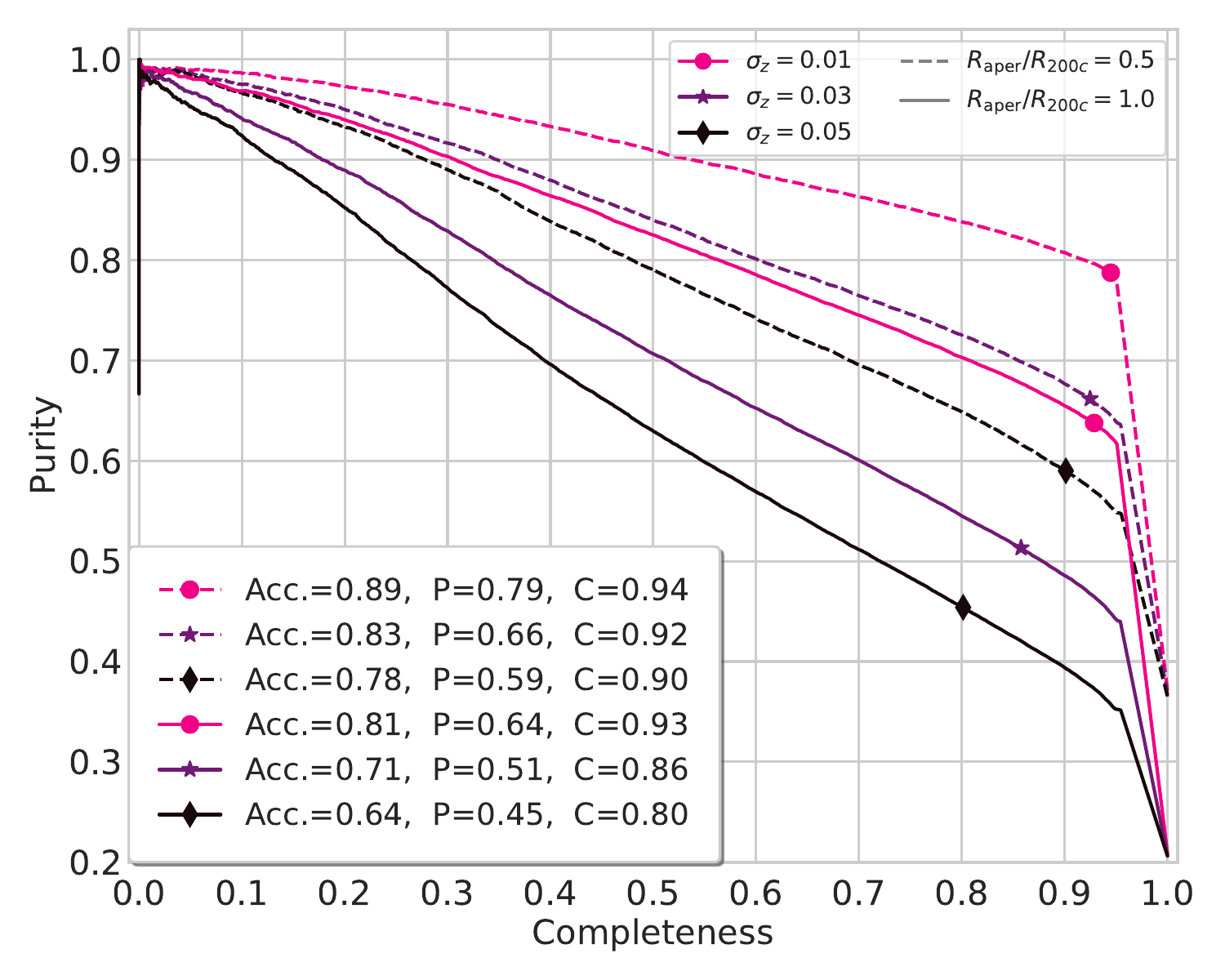}
    \caption{Purity (P) versus completeness (C) for different photo-z samples. The purity of the membership selection is dependent on the accuracy of the photometric redshifts. For instance, with an optimistic accuracy of $\sigma_{z,0} = 0.01$, the purity and completeness reach  an optimal value of $79\%$ and $94\%$ respectively. The DNF photo-z algorithm has an accuracy that is similar to our realistic $\sigma_{z,0} = 0.03$ scenario. However, it has lower completeness due to outliers, which are not present in the artificial Gaussian photo-z sample.}\label{fig:purity_completeness}
\end{figure}

In \autoref{fig:purity_completeness}, \je{we plot} the purity versus completeness for different values of $\sigma_{z,0}$, the photometric redshift uncertainty. The figure \je{ is constructed by varying the cluster membership probability threshold.} The optimal choice \je{(coloured points)} is the one that maximizes the product of both quantities. 

Overall, \je{Copacabana performs well for different photo-z samples when compared with other classifiers \citep{George2011, Castignani2016, Lopes2020}. The product of purity (P) and completeness (C) is maximised a $P=64\%$ and $C=93\%$ in an optimistic scenario ($\sigma_{z,0} = 0.01$), and values of $P=45\%$ and $C=80\%$ in the worst scenario ($\sigma_{z,0} = 0.05$).} It is important to note that the completeness is not higher than $96\%$. The $2\sigma$ photo-z threshold we use translates into $\sim 5\%$ of galaxies having $P_{\rm mem}=0$. \je{This threshold avoids outliers that might be present in the color distributions. }

The membership accuracy is higher for the smaller \Ra{} aperture (see the dashed lines in \autoref{fig:purity_completeness}) and an accuracy of $89 \%$ is achieved in the best scenario. For a given science case, for instance, for studies of the red-sequence, Copacabana can provide an excellent membership selection without relying on previous knowledge of the red-sequence. 

\subsection{\ensuremath{\mu_\star}$-M_{200,c}$ Scaling Relation}\label{sec:mass_observable_relation}
\begin{figure}
    \centering
    \includegraphics[width=0.45\textwidth]{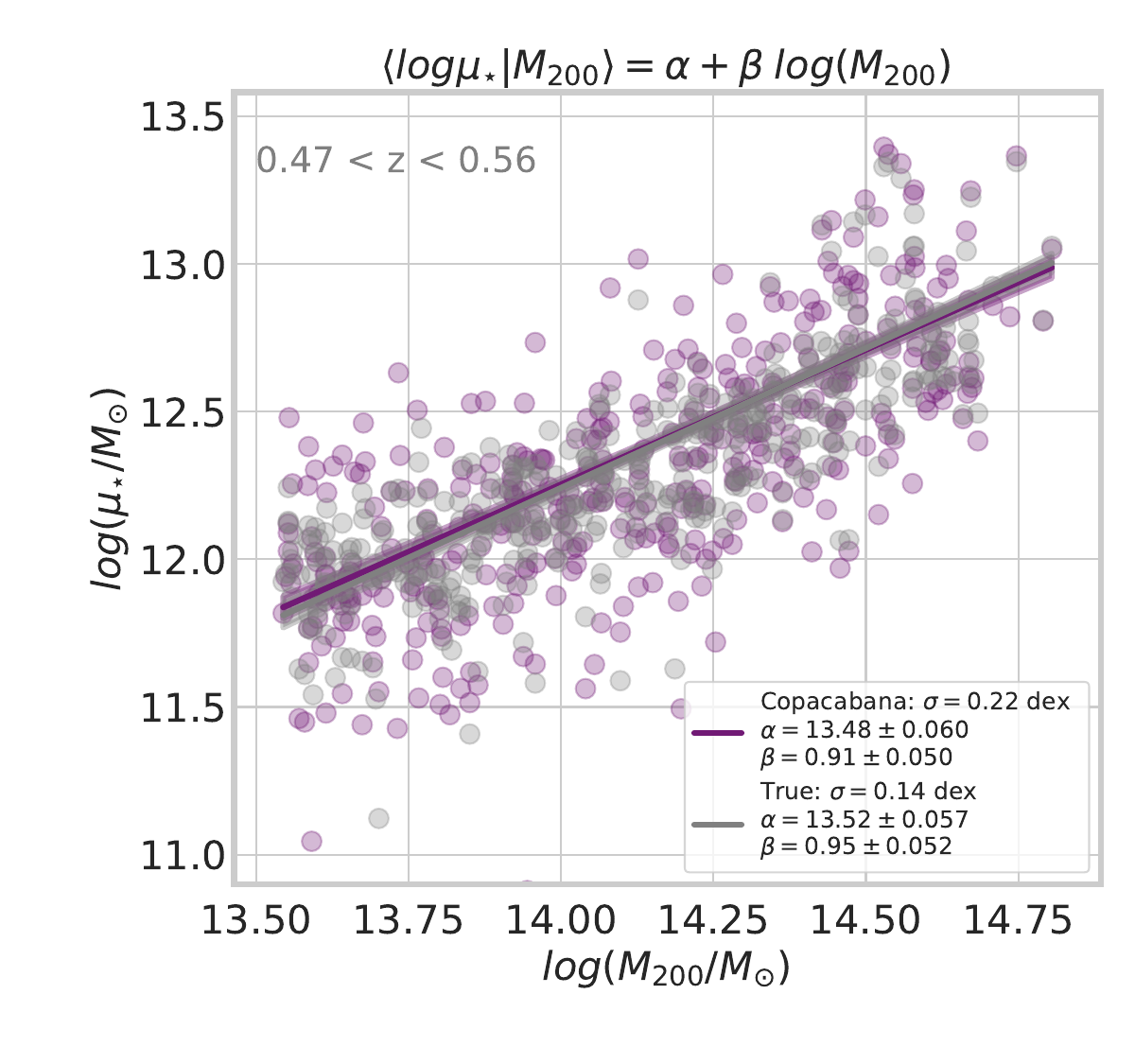}
    \caption{Scaling relation, $\mu_{\star}$ versus $M_{200,c}$. In purple, the DES-like photo-z, $\sigma_{z,0} = 0.03$, and in grey, the true relation, \mustarTrue. The fitted linear relation using a Bayesian regression linear method (linmix) is shown in solid lines, and the $68\%$ confidence level is shown in shaded bands.}
\label{fig:mor_g003}
\end{figure}

\je{An example of the $\mu_{\star}-M_{\rm 200,c}$ scaling relations is shown in \autoref{fig:mor_g003}, using \mustar{} and \mustarTrue{} as the mass proxy for clusters in the the lowest redshift bin, $0.47 < z < 0.56$, and using DES-like photo-z accuracies, $\sigma_{z,0} = 0.03$.} The actual relation, \mustarTrue{}$-M_{\rm 200,c}$ (grey line) is consistent (within $2\sigma$) of with the purple line that Copacabana estimated. We note some small differences between the two curves. For example, the intrinsic scatter is larger, and the slope is shallower for the purple line. These two differences hint at how membership probabilities bias our results of the actual scaling relation. In the following section, we present and discuss the impact of the quality of the photometric redshifts and size cluster aperture on the fitted parameters. 


\subsubsection{Fitted Parameters}

For cosmological parameter estimation, the scatter at a fixed \mustar{} is the important quantity describing the halo mass function. 
At first order, the scatter at fixed mass-proxy ($\sigma_{\log M | \log \mu_{\star}}$) can be written as \citep[e.g.][]{Evrard2014}:
\begin{equation}\label{eq:scatter_fixed_mass}
    \sigma_{\log M | \log \mu_{\star}} = \sigma_{\log{M_{\rm 200,c}}}/ \beta\; .
\end{equation}

A steeper slope results in a lower $\sigma_{\log M | \log \mu_{\star}}$ just as much as a smaller intrinsic scatter $\sigma$. For this reason, we focus on the slope $\beta$ and the scatter $\sigma$ since they are the important parameters for cluster cosmological analysis. 

\begin{figure*}
    \centering
    \includegraphics[width=1.0\textwidth]{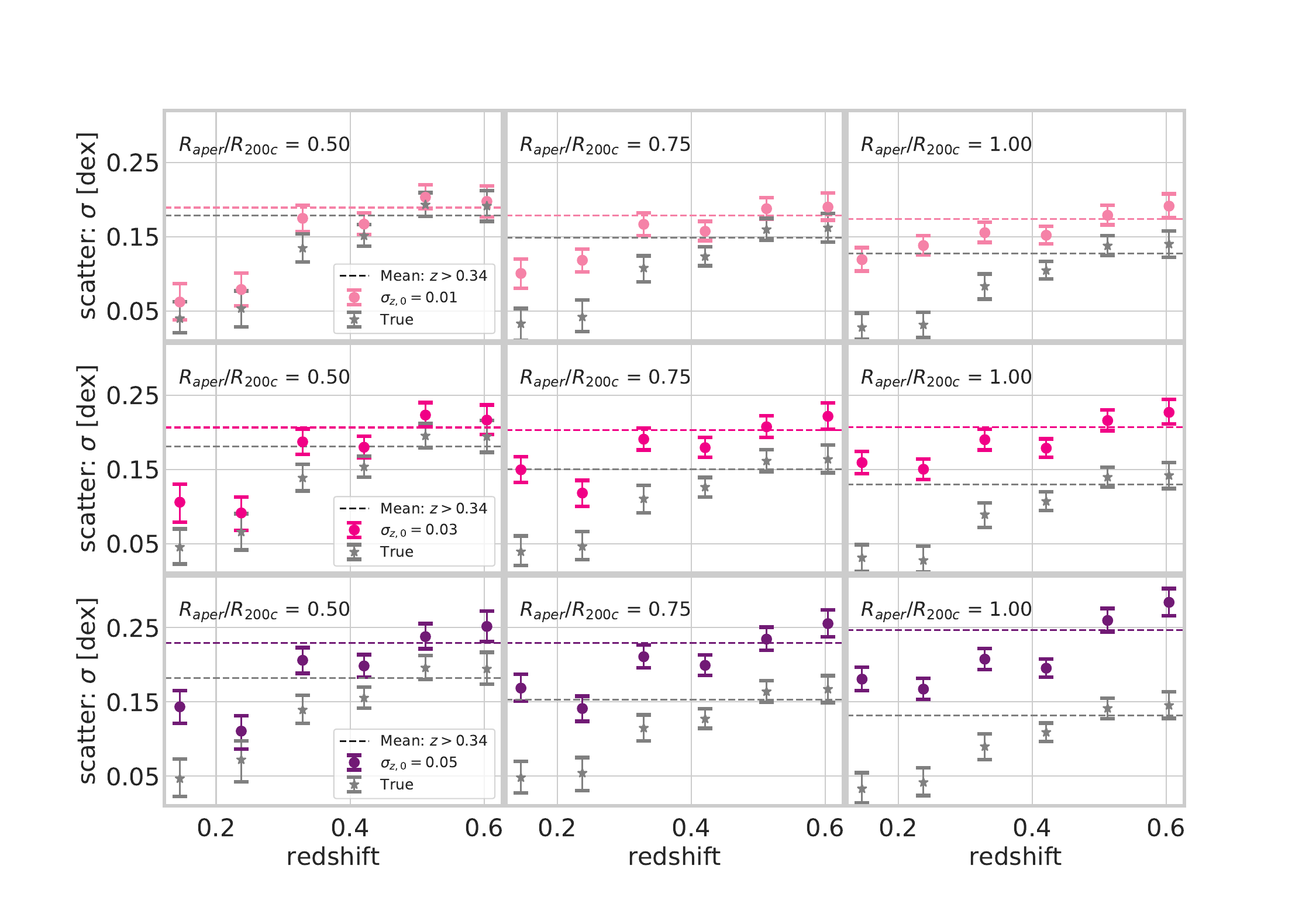}
    \caption{Scatter versus redshift for two stellar mass estimations \mustarTrue{} (True, grey points) and \mustar{}(Copacabana, coloured points). Each row displays the results for a given photo-z sample and each column for a given cluster aperture. Down in columns, the scatter increases as the uncertainty in the photometric redshifts increases, while the scatter for the true relation is fixed. The coloured dashed lines show the mean values of the scatter in each sub-plot for visual comparison with the mean scatter of the true relation (grey dashed lines). The grey stars are the scatter of the true relation, and therefore do not change between rows. Across a row, the scatter in the true relation decreases.}
\label{fig:mor_scatter}
\end{figure*}

\begin{figure*}
    \centering
    \includegraphics[width=1.0\textwidth]{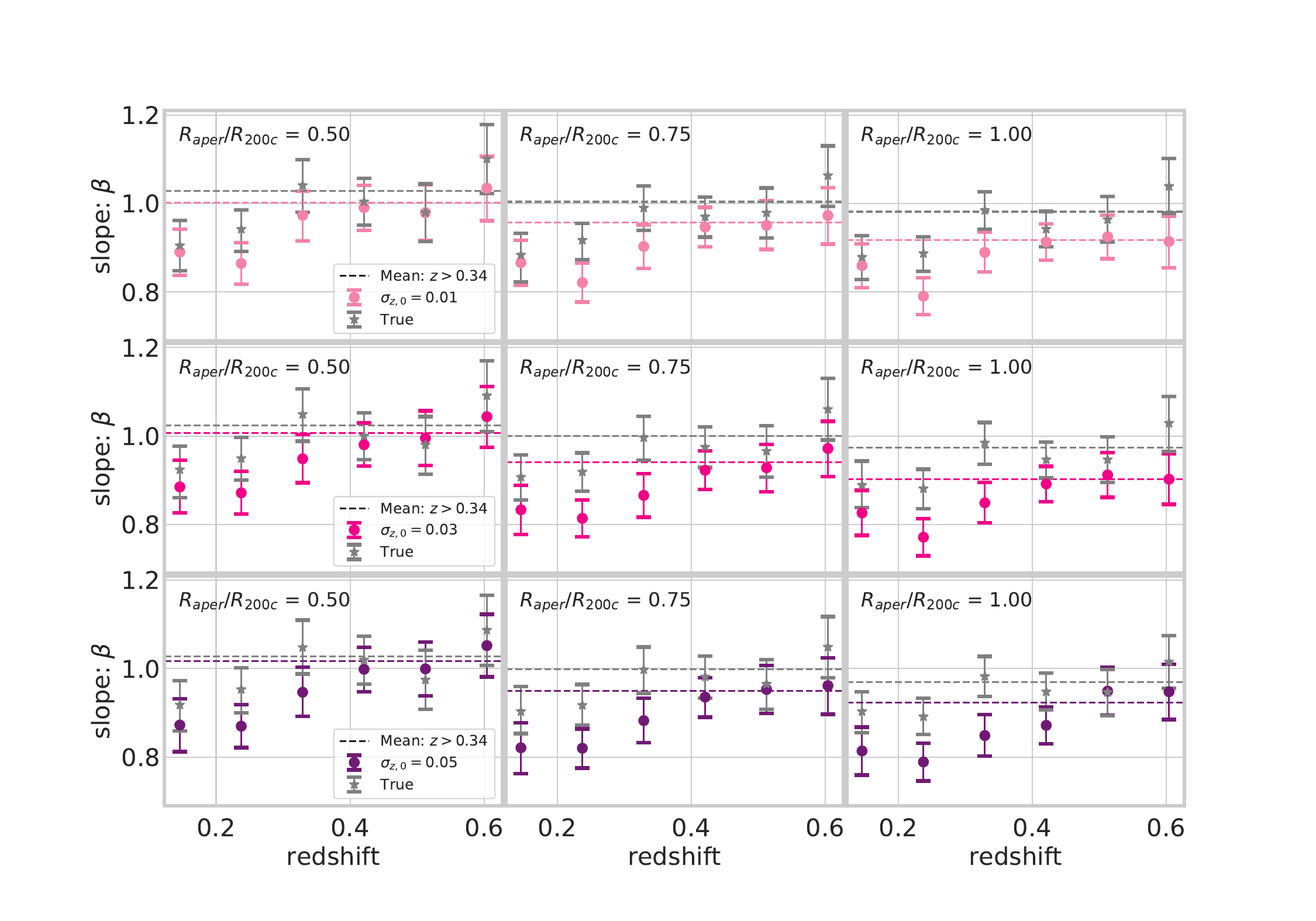}
    \caption{
    As in \autoref{fig:mor_scatter} but for the slope of the scaling relation. A larger cluster aperture increases the contribution from background galaxies and produces a shallower slope. As a result, the \Ra{} aperture has the steepest slope, and it is the only case where the slope derived by Copacabana agrees with the true one irrespective of the photo-z sample used.}
\label{fig:mor_slope}
\end{figure*}

The fitted scatter and slope values as a function of redshift are shown in \autoref{fig:mor_scatter} and \autoref{fig:mor_slope}, respectively. The panel displays three different photo-z precisions across rows and three different cluster apertures across columns. There is an overall shift of the Copacabana from the true values, 
indicated by the mean values (dashed lines). The gap, i.e. the additional shift, increases with a poorer photo-z precision and a larger cluster aperture. 

In particular, the fitted scatter does not change significantly with the cluster aperture. In \autoref{fig:mor_scatter}, the mean values
are about the same across the columns for any photo-z. While the difference between dashed lines is larger, the actual intrinsic scatter (grey stars) decreases with the cluster aperture, which counterbalances the noise added by a larger aperture. The outcome, stressed by the mean value, is that the intrinsic scatter does not depend significantly on the cluster aperture. In contrast, intrinsic scatter did depend on aperture when considering the \mustar{} precision in   \autoref{subsec:mu-star-precision-photoz-cluster-aperture}. 

Regarding the slope, \autoref{fig:mor_slope} demonstrates that a decrease in the accuracy of membership probability tends to result in a shallower slope. As discussed earlier in \autoref{subsec:mu-star-precision-photoz-cluster-aperture}, larger photo-z precision and a larger aperture size can reduce the accuracy of the membership probability. This effect is most noticeable in the bottom left panel of \autoref{fig:mor_slope}, where the most significant discrepancy with the actual value is seen due to the combination of low photo-z precision and a large aperture. When we consider a specific survey scenario (represented by a given row), the mean slope (colorful dashed line) tends to be shallower when the cluster aperture is larger. This trend is reduced with better photo-z accuracy.

\begin{figure*}
    \centering
    \includegraphics[width=1.0\textwidth]{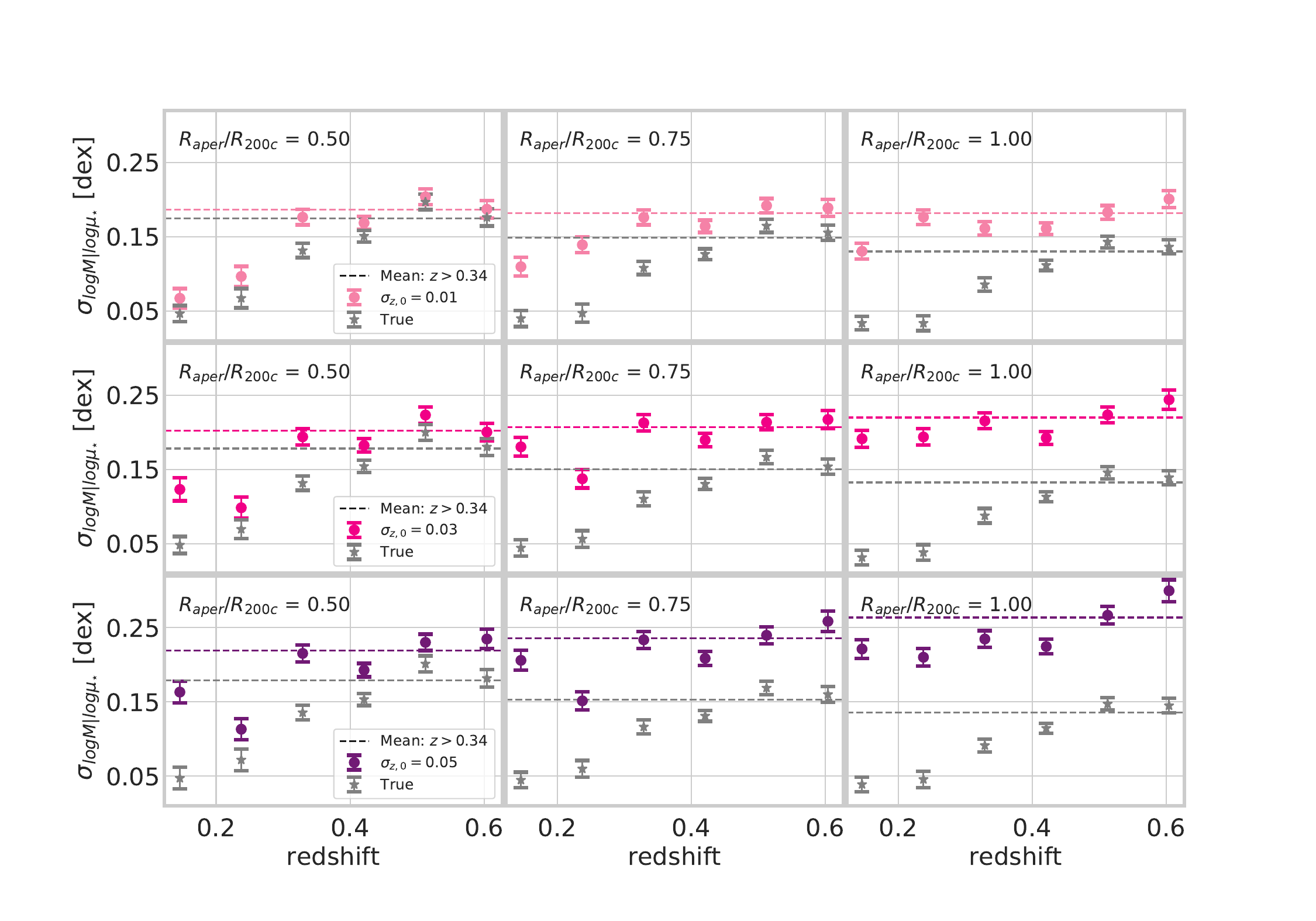}
    \caption{Scatter at fixed observable relation versus the redshift for two independent variables \mustarTrue (True) and \mustar(Copacabana). Each row displays a result for a given photo-z sample, and each column for a given cluster aperture. The photo-z precision impacts the scatter of the recovered \mustar{}$-M_{\rm 200,c}$ scaling relation for any given cluster aperture. The scatter at fixed \mustar{} has a trend with the cluster aperture. The gap between the colourful and grey dashed lines increases across the rows and columns. Note: the grey stars are fixed between rows. }
\label{fig:mor_scatter_fixed_obs}
\end{figure*}

The simultaneous change on the slope and the scatter imply that the scatter at fixed \mustar{}, $\sigma_{\rm log M| log \mu_{\star}}$, is affected by the uncertainty on the membership probabilities. Using \autoref{eq:scatter_fixed_mass}, we can infer $\sigma_{\rm log M| log \mu_{\star}}$ and quantify the impact of our methodology on the scaling relation parameters. The \autoref{fig:mor_scatter_fixed_obs} shows these results displayed similarly to that of \autoref{fig:mor_scatter}.  

A smaller cluster aperture improves the \mustar{} measurement by introducing less noise to the intrinsic scatter of the observable-mass relation. However, for an optimistic photo-z sample such as LSST with $\sigma_{z,0}=0.01$, using scatter computed at the \Rcritical{} is a feasible option, as suggested by the observed trend in the mean scatter value. Conversely, when dealing with an SDSS-like photo-z sample, where $\sigma_{z,0}=0.05$, opting for a smaller cluster aperture may be the optimal approach to reduce the noise introduced by uncertainty in the redshift.

The remarkable precision achieved by using all the cluster population, blue and red galaxies, in photometric data  demonstrates the power of the Copacabana probabilities. In this work, we have obtained a scatter of $0.19$ dex for our mass proxy \mustar{} in a DES-like photo-z sample. For comparison, the typical scatter for $\lambda$ is $0.20 $ dex, and for X-ray, luminosity, and temperature are $0.30$ dex and $0.21$ dex, respectively. Additionally, \mustar{} is an astrophysically motivated mass-proxy since it contains the stellar information of the entire cluster population. Previous works show that the stellar content has a low intrinsic scatter value according to simulations \citep{Anbajagane2020,Ho2023}.
Moreover, there are hints that a mass proxy that only uses the BCG stellar mass has a comparable scatter value to X-ray luminosity and temperature \citep{Mulroy2019,Ho2023}. 

Regarding the redshift evolution of the fitted parameters, we do not see a smooth evolution with redshift as we would expect. 
Instead, there is a jump around redshift $0.3$. This feature is also in the true underlying relation. \citet{DeRose2019arXiv} showed that the switch on the simulation resolution at $z=0.32$ impacts the matter power spectrum of the Buzzard data. This change in resolution might explain the redshift discontinuity seen in this work.

\subsubsection{\mustar\ as a low scatter mass-proxy}
The total galaxy stellar content of halos (\mustarTrue) is a powerful predictor of the cluster halo mass \citep{Behroozi2010, Kravtsov2018, Bradshaw2019arXiv}. Let us consider the case where all the true members inside \Rcritical{} are known and focus on the actual \mustarTrue -- $M_{200,c}$ relation, i.e., without photo-z errors.

Here, we can see the ideal stellar-mass-based mass proxy. The grey points in \autoref{fig:mor_scatter} show the scatter, and the \mustarTrue -- $M_{200,c}$ relation exhibits a low scatter. At the lowest redshift bins, $\sigma$ is about $0.05 \; \text{dex}$, a comparable value to the intra-cluster medium, e.g. $Y_{\rm SZ}$, $M_{\rm gas}$ \citep{Kravtsov2006, Mulroy2019, Bleem2020, Ho2023}, and redshift-based mass-proxies, e.g. velocity dispersions \citep{Ruel2014}. At higher redshifts $z>0.32$, the scatter increases, though we believe this is a simulation effect as the switch of box resolution of the Buzzard simulation occurs at $z=0.33$. 

On the other hand, the cluster aperture significantly affects the true scatter, as in \autoref{fig:mor_scatter}, the mean scatter values (grey dashed lines) decrease over the columns. The best aperture is the \Rc{} as expected by our physical knowledge of halos -- the region that encloses the overall virialized galaxy population should best predict the cluster mass. In contrast, the smaller the aperture weakens the correlation of the visible member stellar masses with the halo mass. 


The stellar mass--halo mass scaling relation from Buzzard is in overall agreement with four hydrodynamical simulations \citep{Anbajagane2020}. In particular, the intrinsic scatter from the Bahamas simulations \citep{Anbajagane2020} at $z=0$ for high mass halos is close to $0.05$ dex, comparable with the lowest redshift result value in  \autoref{fig:mor_scatter_fixed_obs}. On the other hand, the slope has a slightly lower mean value, see \autoref{fig:mor_slope}. Furthermore, a lower slope value was also reported for richness in simulations compared to the redMaPPer DES Y1 measured relations \citep{DeRose2019arXiv}.

\section{Conclusions}\label{sec:conclusions}
This work presents the Copacabana algorithm, which assigns membership probabilities to galaxies in a given cluster field. \je{We validated the algorithm} using the Buzzard simulation. In particular, \je{the stellar mass of cluster galaxies weighted by the membership probabilities, \mustar{}, was found to have an accuracy up to $0.06$ dex for photometric redshifts that have photo-z uncertainties comparable to that expected in LSST data.} \je{In addition, we show that our methodology could precisely recover the scaling relation between \mustar{} and cluster mass, indicating that \mustar{} can be a competitive mass-proxy for optically selected clusters in future cosmological analysis.}

\begin{itemize}
    \item  \textit{\textbf{Performance:}} we show \je{the uncertainty in \mustar{} is mainly affected by the quality of the photometric redshifts}, followed by the cluster aperture. In the best case, \je{photometric redshifts with LSST-like accuracies ($\sigma_{z,0} =0.01 $)}, we reported a \mustar{} uncertainty of $0.09$ dex. A smaller aperture, for instance, \Ra, \je{leads to a smaller uncertainty $\sim 0.05$ dex.}

    \item \textit{\textbf{Cluster Size:}} We present a new method to measure cluster size,  \Rcritical, with only photometric data. The procedure is based on the HOD relation. We inferred a precision of $30\%$ in the context of the Buzzard simulations. \je{The accuracy of our estimate does not depend on the quality of the photometric redshifts or the halo mass.}

    \item \textit{\textbf{Impact on halo mass estimations:}} We quantify how the \mustar{} uncertainty \je{propagates to estimates of the halo mass.} Specifically, we study the parameters of the scaling relation $\mu_{\star}-M_{200c}$ with a focus on the slope and the intrinsic scatter. \je{The photo-z uncertainty} is the main parameter \je{in affecting the scaling relation parameters}. For instance, in the LSST-like scenario, we recover the parameters with no significant difference compared to the true relation. In contrast, \je{in the scenario of large photo-z uncertainties and large apertures, there was a significant impact on recovering the true parameters.} 

    \item \textit{\textbf{How the cluster aperture impacts the halo mass:}}
    While the accuracy of \je{the photometric redshifts is survey-dependent, the aperture size can be adjusted to suit the scientific objectives.} For example, a smaller cluster aperture can significantly improve the membership probabilities. However, that improvement does not necessarily translate into gains in predicting the halo mass. For instance, we do not find a substantial improvement in the scatter of the scaling relation, and the precision gained by a smaller aperture is counterbalanced by the larger intrinsic scatter in $\mu_{\star}-M_{200c}$, which is minimal at \Rcritical. On the other hand, we find that the aperture size significantly affects the slope. In particular, at an aperture of \Ra{}, the recovered slope is very close \je{to the true slope and is insensitive to the range of photometric redshift uncertainties that one encounters in modern photometric surveys.} While at \Rcritical, the recovered slope can be up to $20\%$ shallower. 
   
    \item \textit{\textbf{Membership probability performance:}} We report our galaxy member selection in terms of Purity (P) and Completeness (C). In our best scenario, \je{the accuracy was $81\%$ with P and C of $64\%$ and $93\%$; when we consider a smaller aperture (\Ra{}), these values were $89\%$, $79\%$, and $94\%$, respectively. The membership probabilities substantially improved with a smaller cluster aperture. }
    
\end{itemize}

In sum, Copacabana is a powerful tool to predict the total stellar mass content of galaxies in clusters. \je{In future work, we will apply the stellar mass-proxy \mustar{} in a cosmological analysis using optical data. }



 \section*{Acknowledgements}
% ...
JHE and MSS are funded by the U.S. Department of Energy (DOE) under grant No. DE-SC0019193. MESP is funded by the Deutsche Forschungsgemeinschaft (DFG, German Research Foundation) under Germany’s Excellence Strategy – EXC 2121 ``Quantum Universe'' – 390833306. 

This research uses resources of the National Energy Research Scientific Computing Center (NERSC), a U.S. Department of Energy Office of Science User Facility located at Lawrence Berkeley National Laboratory. 

Funding for the DES Projects has been provided by the U.S. Department of Energy, the U.S. National Science Foundation, the Ministry of Science and Education of Spain, 
the Science and Technology Facilities Council of the United Kingdom, the Higher Education Funding Council for England, the National Center for Supercomputing 
Applications at the University of Illinois at Urbana-Champaign, the Kavli Institute of Cosmological Physics at the University of Chicago, 
the Center for Cosmology and Astro-Particle Physics at the Ohio State University,
the Mitchell Institute for Fundamental Physics and Astronomy at Texas A\&M University, Financiadora de Estudos e Projetos, 
Funda{\c c}{\~a}o Carlos Chagas Filho de Amparo {\`a} Pesquisa do Estado do Rio de Janeiro, Conselho Nacional de Desenvolvimento Cient{\'i}fico e Tecnol{\'o}gico and 
the Minist{\'e}rio da Ci{\^e}ncia, Tecnologia e Inova{\c c}{\~a}o, the Deutsche Forschungsgemeinschaft and the Collaborating Institutions in the Dark Energy Survey. 

The Collaborating Institutions are Argonne National Laboratory, the University of California at Santa Cruz, the University of Cambridge, Centro de Investigaciones Energ{\'e}ticas, 
Medioambientales y Tecnol{\'o}gicas-Madrid, the University of Chicago, University College London, the DES-Brazil Consortium, the University of Edinburgh, 
the Eidgen{\"o}ssische Technische Hochschule (ETH) Z{\"u}rich, 
Fermi National Accelerator Laboratory, the University of Illinois at Urbana-Champaign, the Institut de Ci{\`e}ncies de l'Espai (IEEC/CSIC), 
the Institut de F{\'i}sica d'Altes Energies, Lawrence Berkeley National Laboratory, the Ludwig-Maximilians Universit{\"a}t M{\"u}nchen and the associated Excellence Cluster Universe, 
the University of Michigan, NSF's NOIRLab, the University of Nottingham, The Ohio State University, the University of Pennsylvania, the University of Portsmouth, 
SLAC National Accelerator Laboratory, Stanford University, the University of Sussex, Texas A\&M University, and the OzDES Membership Consortium.

Based in part on observations at Cerro Tololo Inter-American Observatory at NSF's NOIRLab (NOIRLab Prop. ID 2012B-0001; PI: J. Frieman), which is managed by the Association of Universities for Research in Astronomy (AURA) under a cooperative agreement with the National Science Foundation.

The DES data management system is supported by the National Science Foundation under Grant Numbers AST-1138766 and AST-1536171.
The DES participants from Spanish institutions are partially supported by MICINN under grants ESP2017-89838, PGC2018-094773, PGC2018-102021, SEV-2016-0588, SEV-2016-0597, and MDM-2015-0509, some of which include ERDF funds from the European Union. IFAE is partially funded by the CERCA program of the Generalitat de Catalunya.
Research leading to these results has received funding from the European Research
Council under the European Union's Seventh Framework Program (FP7/2007-2013) including ERC grant agreements 240672, 291329, and 306478.
We  acknowledge support from the Brazilian Instituto Nacional de Ci\^encia
e Tecnologia (INCT) do e-Universo (CNPq grant 465376/2014-2).

This manuscript has been authored by Fermi Research Alliance, LLC under Contract No. DE-AC02-07CH11359 with the U.S. Department of Energy, Office of Science, Office of High Energy Physics.

\section*{Data Availability}
The data used in this work is available upon request.




\bibliographystyle{mnras}
\bibliography{bib} 



\appendix

\section{Affiliations}

\scriptsize
$^{1}$ Department of Physics, University of Michigan, Ann Arbor, MI 48109, USA\\
$^{2}$ Hamburger Sternwarte, Universit\"{a}t Hamburg, Gojenbergsweg 112, 21029 Hamburg, Germany\\
$^{3}$ Fermi National Accelerator Laboratory, P. O. Box 500, Batavia, IL 60510, USA\\
$^{4}$ Departments of Statistics and Data Science, University of Texas at Austin, Austin, TX 78757, USA\\
$^{5}$ \\
$^{6}$ Department of Physics, Carnegie Mellon University, Pittsburgh, Pennsylvania 15312, USA\\
$^{7}$ Observational Cosmology Lab, NASA Goddard Space Flight Center, Greenbelt, MD 20771, USA\\
$^{8}$ Department of Astronomy, University of Maryland, College Park, 20742, USA\\
$^{9}$ Department of Physics, Boise State University, Boise, ID 83725, USA\\
$^{10}$ Laborat\'orio Interinstitucional de e-Astronomia - LIneA, Rua Gal. Jos\'e Cristino 77, Rio de Janeiro, RJ - 20921-400, Brazil\\
$^{11}$ Institute of Cosmology and Gravitation, University of Portsmouth, Portsmouth, PO1 3FX, UK\\
$^{12}$ University Observatory, Faculty of Physics, Ludwig-Maximilians-Universit\"at, Scheinerstr. 1, 81679 Munich, Germany\\
$^{13}$ Department of Physics \& Astronomy, University College London, Gower Street, London, WC1E 6BT, UK\\
$^{14}$ Instituto de Astrofisica de Canarias, E-38205 La Laguna, Tenerife, Spain\\
$^{15}$ Universidad de La Laguna, Dpto. Astrofísica, E-38206 La Laguna, Tenerife, Spain\\
$^{16}$ Institut de F\'{\i}sica d'Altes Energies (IFAE), The Barcelona Institute of Science and Technology, Campus UAB, 08193 Bellaterra (Barcelona) Spain\\
$^{17}$ Astronomy Unit, Department of Physics, University of Trieste, via Tiepolo 11, I-34131 Trieste, Italy\\
$^{18}$ INAF-Osservatorio Astronomico di Trieste, via G. B. Tiepolo 11, I-34143 Trieste, Italy\\
$^{19}$ Institute for Fundamental Physics of the Universe, Via Beirut 2, 34014 Trieste, Italy\\
$^{20}$ Centro de Investigaciones Energ\'eticas, Medioambientales y Tecnol\'ogicas (CIEMAT), Madrid, Spain\\
$^{21}$ Jet Propulsion Laboratory, California Institute of Technology, 4800 Oak Grove Dr., Pasadena, CA 91109, USA\\
$^{22}$ Kavli Institute for Cosmological Physics, University of Chicago, Chicago, IL 60637, USA\\
$^{23}$ Instituto de Fisica Teorica UAM/CSIC, Universidad Autonoma de Madrid, 28049 Madrid, Spain\\
$^{24}$ Center for Astrophysical Surveys, National Center for Supercomputing Applications, 1205 West Clark St., Urbana, IL 61801, USA\\
$^{25}$ Department of Astronomy, University of Illinois at Urbana-Champaign, 1002 W. Green Street, Urbana, IL 61801, USA\\
$^{26}$ School of Mathematics and Physics, University of Queensland,  Brisbane, QLD 4072, Australia\\
$^{27}$ Santa Cruz Institute for Particle Physics, Santa Cruz, CA 95064, USA\\
$^{28}$ Center for Cosmology and Astro-Particle Physics, The Ohio State University, Columbus, OH 43210, USA\\
$^{29}$ Department of Physics, The Ohio State University, Columbus, OH 43210, USA\\
$^{30}$ Center for Astrophysics $\vert$ Harvard \& Smithsonian, 60 Garden Street, Cambridge, MA 02138, USA\\
$^{31}$ Australian Astronomical Optics, Macquarie University, North Ryde, NSW 2113, Australia\\
$^{32}$ Lowell Observatory, 1400 Mars Hill Rd, Flagstaff, AZ 86001, USA\\
$^{33}$ Centre for Gravitational Astrophysics, College of Science, The Australian National University, ACT 2601, Australia\\
$^{34}$ The Research School of Astronomy and Astrophysics, Australian National University, ACT 2601, Australia\\
$^{35}$ Departamento de F\'isica Matem\'atica, Instituto de F\'isica, Universidade de S\~ao Paulo, CP 66318, S\~ao Paulo, SP, 05314-970, Brazil\\
$^{36}$ George P. and Cynthia Woods Mitchell Institute for Fundamental Physics and Astronomy, and Department of Physics and Astronomy, Texas A\&M University, College Station, TX 77843,  USA\\
$^{37}$ LPSC Grenoble - 53, Avenue des Martyrs 38026 Grenoble, France\\
$^{38}$ Instituci\'o Catalana de Recerca i Estudis Avan\c{c}ats, E-08010 Barcelona, Spain\\
$^{39}$ Department of Astrophysical Sciences, Princeton University, Peyton Hall, Princeton, NJ 08544, USA\\
$^{40}$ Observat\'orio Nacional, Rua Gal. Jos\'e Cristino 77, Rio de Janeiro, RJ - 20921-400, Brazil\\
$^{41}$ Kavli Institute for Particle Astrophysics \& Cosmology, P. O. Box 2450, Stanford University, Stanford, CA 94305, USA\\
$^{42}$ SLAC National Accelerator Laboratory, Menlo Park, CA 94025, USA\\
$^{43}$ Department of Physics and Astronomy, Pevensey Building, University of Sussex, Brighton, BN1 9QH, UK\\
$^{44}$ Instituto de F\'\i sica, UFRGS, Caixa Postal 15051, Porto Alegre, RS - 91501-970, Brazil\\
$^{45}$ School of Physics and Astronomy, University of Southampton,  Southampton, SO17 1BJ, UK\\
$^{46}$ Computer Science and Mathematics Division, Oak Ridge National Laboratory, Oak Ridge, TN 37831\\
$^{47}$ Lawrence Berkeley National Laboratory, 1 Cyclotron Road, Berkeley, CA 94720, USA\\
$^{48}$ Department of Physics, Duke University Durham, NC 27708, USA\\

\bsp	
\label{lastpage}
\end{document}